\documentclass[conference]{IEEEtran}
\IEEEoverridecommandlockouts
\PassOptionsToPackage{table}{xcolor}
\usepackage{amsmath,amssymb,amsfonts}
\usepackage{algorithmic}
\usepackage{graphicx}
\usepackage{textcomp}
\usepackage{hyperref}
\usepackage{algorithm}
\usepackage{multirow}
\usepackage{float}
\usepackage{url}
\usepackage{hyperref}
\usepackage{amssymb}
\usepackage{comment}
\usepackage{xspace,framed}
\usepackage{siunitx}
\usepackage{times}
\usepackage[inline]{enumitem}
\usepackage{listings}
\usepackage{cite}
\newcommand\tool{\textit{EBF}\xspace}
\usepackage{soul}
\usepackage{llvm}
\usepackage{llvmstyle}
\usepackage[mode=buildnew]{standalone}
\usepackage[caption=false]{subfig}
\usepackage{caption,setspace}
\usepackage[normalem]{ulem}

\newcolumntype{"}{@{\hskip\tabcolsep\vrule width 1.5pt\hskip\tabcolsep}}
\makeatother

\def\BibTeX{{\rm B\kern-.05em{\sc i\kern-.025em b}\kern-.08em
    T\kern-.1667em\lower.7ex\hbox{E}\kern-.125emX}}
\begin{document}

\title{Combining BMC and Fuzzing Techniques for Finding Software Vulnerabilities in Concurrent Programs}

\author{
\IEEEauthorblockN {
   Fatimah K. Aljaafari$^{1,2}$,       
   Rafael Menezes$^{1,3}$       
   Edoardo Manino$^{1}$       
   Fedor Shmarov$^{1}$  \\     
   Mustafa A. Mustafa$^{1,4}$       
   and Lucas C. Cordeiro$^{1,3}$ 
}

\IEEEauthorblockA {
  $^{1}$Department of Computer Science\\
  The University of Manchester\\
  Manchester, UK \\
  $^{2}${King Faisal University, Al-Hofuf, Saudi Arabia}\\
    $^{3}${Federal University of Amazonas, Manaus, Brazil}
}
\IEEEauthorblockA {
    $^{4}$imec-COSIC, KU Leuven, Leuven, Belgium
}
}

\markboth
{Aljaafari \headeretal: Combining BMC and Fuzzing Techniques for Finding Software Vulnerabilities in Concurrent Programs }
{Aljaafari \headeretal: Combining BMC and Fuzzing Techniques for Finding Software Vulnerabilities in Concurrent Programs}

\maketitle

\maketitle

\begin{abstract}
Finding software vulnerabilities in concurrent programs is a challenging task due to the size of the state-space exploration, as the number of interleavings grows exponentially with the number of program threads and statements. We propose and evaluate \tool (Ensembles of Bounded Model Checking with Fuzzing) -- a technique that combines Bounded Model Checking (BMC) and Gray-Box Fuzzing (GBF) to find software vulnerabilities in concurrent programs. Since there are no publicly-available GBF tools for concurrent code, we first propose \textit{OpenGBF} -- a new open-source concurrency-aware gray-box fuzzer that explores different thread schedules by instrumenting the code under test with random delays. Then, we build an ensemble of a BMC tool and \textit{OpenGBF} in the following way. On the one hand, when the BMC tool in the ensemble returns a counterexample, we use it as a seed for \textit{OpenGBF}, thus increasing the likelihood of executing paths guarded by complex mathematical expressions. On the other hand, we aggregate the outcomes of the BMC and GBF tools in the ensemble using a decision matrix, thus improving the accuracy of \tool. We evaluate \tool against state-of-the-art pure BMC tools and show that it can generate up to $14.9\%$ more correct verification witnesses than the corresponding BMC tools alone. Furthermore, we demonstrate the efficacy of \textit{OpenGBF}, by showing that it can find $24.2\%$ of the vulnerabilities in our evaluation suite, while non-concurrency-aware GBF tools can only find $0.55\%$. Finally, thanks to our concurrency-aware \textit{OpenGBF}, \tool detects a data race in the open-source \textit{wolfMqtt} library and reproduces known bugs in several other real-world programs, which demonstrates its effectiveness in finding vulnerabilities in real-world software. 
\end{abstract}


\section{Introduction}
\label{sec:introduction}

Concurrency is becoming prevalent in present-day software systems thanks to the performance benefits provided by multi-core hardware~\cite{sodan2010parallelism}. 
Examples of such software systems include online banking, auto-pilots, computer games and railway ticket reservation systems~\cite{multithreadExample}.
Ensuring the correctness and safety of such software is crucial since software failures may lead to significant financial losses and affect people's well-being~\cite{CordeiroFB20}. As an example, the OpenSSL library had a Heartbleed\footnote{https://heartbleed.com/} vulnerability that allows a remote attacker to get access to sensitive information.

Despite the significant resources invested into software testing, much of existing software still features security vulnerabilities~\cite{lu2008learning} 
This is because the different possible threads' interleavings cause the program execution to be non-deterministic, thus making the process of testing and verifying concurrent programs an inherently difficult task~\cite{PereiraASMMFC17} 
(e.g., some bugs may occur only for a specific threads order, making them harder to detect). 
Furthermore, there exists a wide variety of unwanted concurrent behaviors. On the one hand, the non-determinism of the thread interleavings introduces concurrency bugs such as data races, deadlocks, thread leaks, and resource starvation~\cite{lu2008learning}, which may cause the program to produce abnormal results or unforeseen hangs. 
On the other hand, specific program inputs and thread interleavings may lead to memory corruption and security violations (e.g., access out of bounds)~\cite{MonteiroASICF18}.

Due to this complexity, manual testing of concurrent software is not always adequate, and so automated verification and testing are often employed.
In this respect, there is a myriad of different automated techniques such as control engineering~\cite{controlEngeneering}, abstract interpretation~\cite{AbstractInterpretation} and data-flow analysis~\cite{dwyer1994data} for detecting bugs and vulnerabilities in concurrent programs~\cite{16,50}. Among those, two methods have seen significant development in recent years: Bounded Model Checking (BMC) and fuzzing~\cite{26}.

BMC~\cite{Biere09} searches for violations in bounded executions (up to some given depth $k$) of the given program. If no property violation is detected, then $k$ is increased until a bug is found, the verification problem becomes intractable, or a pre-set upper bound is reached. Although many industrial-grade bounded model checkers~\cite{gadelha2020esbmc, kroening2014cbmc, cpachecker2011, he2021satisfiability, 95} have been successfully used for software verification,
BMC has several fundamental drawbacks in general. Namely, BMC often experiences difficulties with achieving high path coverage (especially for multi-threaded programs) and reaching deep statements within the code because of state-space explosion and its dependency on Boolean Satisfiability (SAT)~\cite{25} or Satisfiability Modulo Theories (SMT) solvers~\cite{69}.

Fuzzing~\cite{50} is an automated software testing technique that involves the repeated generation of inputs (based on some initial guess -- seed value) to a Program Under Test (PUT). The PUT is then executed for each given sequence of input values; its behavior is checked for abnormalities, such as crashes or failures~\cite{27}.
The main advantages of fuzzing include relative ease of integration with the existing testing frameworks, high scalability, and most importantly, exploring the deep execution paths is not as costly as in BMC.
However, fuzzing often suffers from low branch coverage since the input generation is based on random mutations~\cite{51}. Typically this occurs when a program features conditional statements with complex conditions (e.g., input validation functions). As a result, providing a good initial seed for the fuzzing process is crucial. 
Moreover, fuzzing techniques face challenges detecting vulnerabilities in multi-threaded programs~\cite{87} since existing fuzzing techniques do not focus on thread interleavings that affect execution states.

Efforts toward developing a combined verification technique harnessing the strengths of both BMC and fuzzing have been made in the past. For example, Ognawala et al.~\cite{28} combine symbolic execution and fuzzing and apply it to general-purpose software. Alshmrany et al.~\cite{alshmrany2020fusebmc} use BMC to guide a fuzzer in the analysis of sequential C programs. Chowdhury et al.~\cite{chowdhury2019verifuzz} improve the seeding of gray-box fuzzing (GBF) by using BMC as a constraint solver to find execution paths through complex blocks of code. Nevertheless, given the current knowledge in software verification, there are no techniques that harness both BMC and fuzzing for verification of concurrent programs, and the question of whether \textit{combining BMC and fuzzing improves bug finding in concurrent programs} remains open.

The challenge in answering this fundamental question is twofold. First, while there are many available BMC tools in the literature, all existing concurrency fuzzers are (at least partially) closed source. As a result, employing any of these concurrency fuzzers requires a major reproducibility effort. Second, combining BMC and fuzzing for concurrency is not straightforward. Given the lack of existing baselines, we take inspiration from \textit{portfolios} \cite{XuSAT2007,BeyerCombo2022}, the practice of running an ensemble of similar tools in parallel and picking the best result. At the same time, BMC and fuzzing are very dissimilar approaches, thus their cooperation inside the ensemble has to be carefully coordinated.

This paper addresses these challenges and makes the following original contributions:
\begin{enumerate}
    \item We develop \textit{OpenGBF} -- a new open-source state-of-the-art concurrency-aware gray-box fuzzer \cite{EBF}. Our main technique is instrumenting the PUT with random delays obtained from a random number generator whose seed value is controlled by the fuzzer. In this way, we can discover different thread interleavings and explore deep execution paths. Furthermore, our fuzzer is capable of generating crash reports containing the full program execution path.
    \item We introduce \tool ~-- Ensembles of Bounded Model Checking with Fuzzing. This technique combines the strengths of BMC in resolving complex conditional guards with the flexibility of our concurrency-aware gray-box fuzzer. \tool incorporates a result decision matrix for coping with the potentially conflicting verdicts produced by the tools in the ensemble. Furthermore, \tool efficiently distributes the available computational resources between the tools to enhance its bug-finding capabilities. 
    \item We demonstrate that the combination of BMC and fuzzing improves verification outcomes compared to either technique applied separately. More specifically, \tool improves the bug-finding abilities of all
    state-of-the-art concurrent BMC tools considered in this work by up to $14.9\%$. Similarly, \tool can find $24.2\%$ of the vulnerabilities in our evaluation suite, whereas the state-of-the-art gray-box fuzzer \textit{AFL++} can only find $0.55\%$. 
    \item We apply \tool to the \textit{wolfMQTT} open-source library that implements the \textit{MQTT} messaging protocol, and we discover the presence of a data race bug. We reported the bug to the developers of the \textit{wolfMQTT} library, who fixed it in June 2021. Also, \textit{EBF} successfully reproduced known bugs in several real-world concurrent programs (i.e., \textit{pfscan}~\cite{pfscan}, \textit{bzip2smp}~\cite{bzip2smp} and \textit{swarm 1.1~}\cite{bader2007swarm}). This demonstrates the real-world capabilities of \tool.
    \item We report that the bug-finding capabilities of \tool are stable across a wide range of parameter values. In detail, we run a comparison experiment along three different axis: time allocation between the BMC tool and \textit{OpenGBF}, maximum delay inserted by \textit{OpenGBF} and maximum number of threads allowed by \textit{OpenGBF}. Our results show a large sweet spot of parameter values that allows \tool to find nearly 50-fold more bugs than the worst setting.
\end{enumerate}

This remaining of the paper is structured as follows: Section~\ref{sec:preliminaries} contains the preliminaries on concurrent programs, bounded model checking and fuzzing, while Section~\ref{sec:probleStatement} states the main research question of this work. Section~\ref{sec:comcurrentFuzzer} discusses the main design choices and implementation details of \textit{OpenGBF}, our state-of-the-art fuzzer for concurrent programs. Section~\ref{sec:compining} presents \tool, our ensemble verification technique. Section~\ref{sec:results} presents the experimental results, while Section~\ref{sec:relatedWork} lists the related work, and Section~\ref{sec:conclusion} draws the final conclusions.

\section{PRELIMINARIES}
\label{sec:preliminaries}

\subsection{Common Software Vulnerabilities}
\label{sec:prelim_concurrency}

Concurrent programs feature multiple processes or threads simultaneously operating on shared computing resources~\cite{PereiraAMSCCSF16}. As a result, such programs can feature vulnerabilities specific to sequential problems (e.g., invalid memory accesses, memory leaks \cite{92}) as well as types of bugs that only occur in concurrent programs (e.g., data races, deadlocks, thread leaks \cite{benari2006}). Some software vulnerabilities are considered more dangerous than others. For example, writing out of bounds (a type of invalid memory access) is ranked number $1$ in the top $25$ MITRE ranking in 2022 \footnote{\url{https://cwe.mitre.org/top25/archive/2022/2022_cwe_top25.html}}, while data races are in the 22nd place.

\textbf{Invalid memory accesses} comprise a vast family of memory safety violations. They include accessing memory outside the bounds of the intended buffer for either reading (potentially revealing some sensitive data to the attacker) or writing (causing memory corruptions or injections of executable code), accessing previously freed memory (aka ``use-after-free''), and dereferencing of invalid pointers or \textit{NULL} pointers (causing the program to crash or exit unexpectedly).

\begin{figure}
    \subfloat[\textbf{Data race} occurs when \texttt{T1} and \texttt{T2} are trying to write to the memory region \texttt{A} simultaneously with no synchronization between the operations. 
    ]{
        \includegraphics[width=0.49\textwidth]{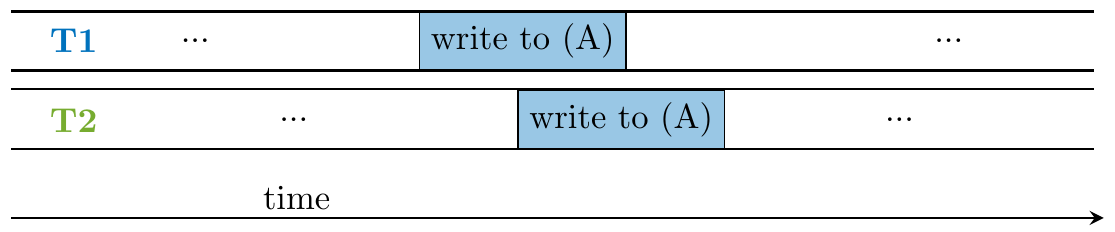}
        \label{fig:data-race}
    } \\
    \subfloat[
    The program ends in a \textbf{deadlock} since \texttt{T1} acquires a lock for the memory region \texttt{A} and then tries to write to the memory region \texttt{B}. At the same time, \texttt{T2} performs the opposite acquiring a lock for \texttt{B} and attempting to write to \texttt{A}. This will result in both threads waiting indefinitely for each other to release their corresponding locks before the program's execution can continue. 
    ]{
        \includegraphics[width=0.49\textwidth]{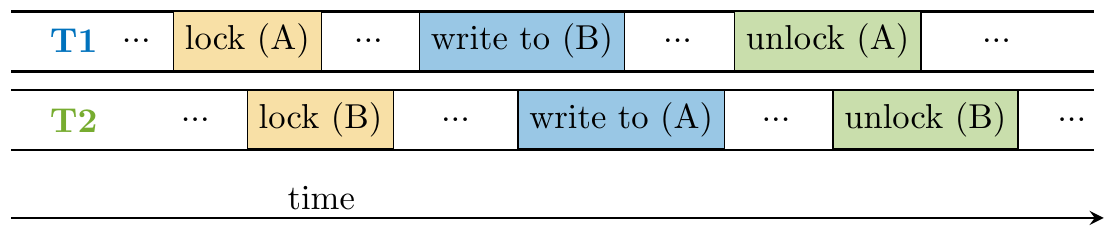}
        \label{fig:deadlock}
    } \\
    \subfloat[\texttt{T3} is a source of \textbf{thread leak} since, unlike \texttt{T2}, it terminates but never joins \texttt{T1}. Therefore, the number of unused threads increases with time causing a potential resource exhaustion.]{
        \includegraphics[width=0.49\textwidth]{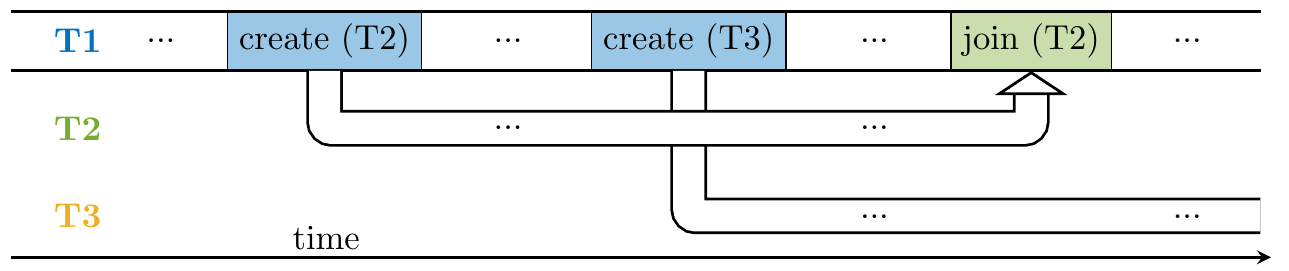}
        \label{fig:thread-leak}
    }
\caption{Concurrency bugs.}
\end{figure}

{\bf Data race} is a condition when the program execution results in an undesired behavior due to a particular sequence and/or timing of the instructions executed by each thread. For instance, when a thread modifies the shared memory without acquiring a lock first, causing memory corruption when another thread tries to update the same memory location (see Figure \ref{fig:data-race}).

{\bf Deadlock} occurs when the program is not in the terminal state and it cannot progress to any other state. For instance, when a thread does not release a lock after accessing the shared memory, therefore, denying memory access for any other thread (see Figure \ref{fig:deadlock}).

{\bf Thread leak} is a vulnerability specific to multi-threaded programs that happens when a terminated thread never joins the calling thread, thus never releasing the occupied resources (see Figure \ref{fig:thread-leak}).

Similarly, \textbf{memory leaks} are caused by repeated memory allocations which are never released during the program's execution. This may lead to memory exhaustion resulting in the system hanging or crashing.

\subsection{Bounded Model Checking}
\label{sec:prelim_bmc}

Bounded model checking is a verification technique that has been successfully applied to software and hardware verification over the past decades \cite{Biere09}.
BMC works with the underlying program's mathematical model (represented as a finite state transition system). It explores the model's evolution up to some finite positive bound $k$ and determines whether the given safety property (e.g., absence of deadlocks, data races, buffer overflows, assertion violations, etc.) holds. In short, BMC symbolically executes the given program up to the given bound $k$ and encodes all the obtained traces $C$ together with the given property $P$ as a SAT/SMT~\cite{satsmt2006} formula $C \wedge \neg P$. A decision procedure (often referred to as \textit{automated theorem prover} or \textit{solver}) then checks the obtained formula and returns the satisfiability verdict. If the formula is satisfiable, then it means that the safety property is violated, and a witness (counter-example) is produced. Otherwise, a proof can be obtained that the program is safe up to the given bound $k$.

Several drawbacks of BMC include state-space explosion as the verification depth grows, which becomes even more challenging for multi-threaded programs since it is required to explore the combined search space of thread interleavings and program states.
Moreover, the verification of logical formulae consumes more CPU time and computer memory as the size of the formulae grows with the increasing verification depth. Finally, since BMC works with a symbolic abstraction (over-approximation) of the underlying program, it may report incorrect results when the devised model does not precisely represent the given program.
For example, this can be caused by external libraries whose implementation in the language supported by the given BMC tool does not exist. Consequently, their behavior must be modelled (approximated) inside the BMC tool.

Thus, existing BMC tools like \textit{ESBMC} \cite{78}, \textit{CBMC} \cite{cbmc} and \textit{Cseq} \cite{Cseq} differ mainly in their choices of program encoding and symbolic abstractions. We provide more details on their strategies to deal with concurrent programs in Section \ref{sec:relatedWork}.

\subsection{Gray-Box Fuzzing}
\label{sec:prelim_fuzzing}

Fuzzing is an automated testing technique that discovers vulnerabilities by repeatedly executing a program with randomly-generated inputs~\cite{AlshmranyABC21}. Since most inputs generated this way are invalid, state-of-the-art fuzzers let users specify a small set of valid program inputs (the seeds) and employ a mutation-based strategy to generate new ones. Gray-box fuzzing improves on this idea by guiding the mutation process with program-specific metrics. To do so, the program under test must be instrumented with some additional code that tracks the required metric (e.g., code coverage) during execution.

\begin{algorithm}
\caption{Gray-Box Fuzzing}\label{alg:AFL}
\textbf{Input:} $PUT$ -- program under test, $M$ -- corpus of initial seeds.\\
\textbf{Output:} $Q_S$ -- seed queue, $S_I$ -- crash inputs found
\begin{algorithmic}[1]
\STATE $P_f\gets instrument(PUT)$  \hfill\COMMENT{instrument the PUT}\label{alg:AFL:line1}
\STATE $Q_S \gets M$ \hfill\COMMENT{initialize the seed queue}
\STATE $S_I = \emptyset$ 
\WHILE{{\bf not} $timeout$}\label{alg:AFL:line4}
\STATE $t \gets select\_next\_seed(Q_S)$ \hfill\COMMENT{pick seed from queue}\label{alg:AFL:line5}
\STATE $N \gets get\_mutation\_chance(P_f, t)$\label{alg:AFL:line6}
\FORALL{$i \in 1 \ldots N$}
    \STATE $t' \gets mutate\_input(t)$ \hfill\COMMENT{mutate the seed}\label{alg:AFL:line8}
    \STATE $rep \gets run(P_f, t', M_c)$ \hfill\COMMENT{execute the PUT}\label{alg:AFL:line9}
    \IF{$is\_crash(rep)$} \label{alg:AFL:line10}
        \STATE $S_I \gets$ $S_I \cup {t'}$\hfill\COMMENT{new vulnerable input found!} \label{alg:AFL:line11}
    \ELSIF{$covers\_new\_trace(t', rep)$}\label{alg:AFL:line12}
        \STATE $Q_S \gets Q_S \oplus t'$ \hfill\COMMENT{add promising seeds to queue}\label{alg:AFL:line13}
    \ENDIF
\ENDFOR
\ENDWHILE
\end{algorithmic}
\end{algorithm}

Algorithm~\ref{alg:AFL}~\cite{lemieux2018fairfuzz,87} shows the standard workflow of a gray-box fuzzer. It takes a target PUT and initial seeds $M$ as inputs. Then, it instruments the PUT (line \ref{alg:AFL:line1}) by inserting some additional code that allows the fuzzer to collect code coverage statistics in the PUT. At every iteration of the main fuzzing loop (line \ref{alg:AFL:line4}), it selects a seed $t$ (line \ref{alg:AFL:line5}) and chooses a random number $N$ of mutations (line \ref{alg:AFL:line6}). Then, the fuzzer repeatedly executes the instrumented program $P_f$ (line \ref{alg:AFL:line9}) with different mutated seed $t'$ (line \ref{alg:AFL:line8}) as input and obtains the execution statistics. If $t'$ triggers a crash in the instrumented program $P_f$ (line \ref{alg:AFL:line10}), it is added to the set of vulnerable inputs (line \ref{alg:AFL:line11}). Otherwise, if $t'$ does not cause a crash but covers a new branch in the PUT (line \ref{alg:AFL:line12}), it is added to the seed queue $Q_S$ (line \ref{alg:AFL:line13}). This may help the fuzzer discover more vulnerabilities in the subsequent iterations. Finally, the execution of the main fuzzing loop continues until the predefined timeout is reached.

\begin{table*}[t]
\centering
    \begin{tabular}{ |c|c|c|c|c|c| } 
    \hline
    Fuzzer Name & \cellcolor{gray!25} Scope & \cellcolor{gray!25} Vulnerabilities & \cellcolor{gray!25} Open Src. & \cellcolor{gray!25} Interleaving Control & \cellcolor{gray!25} Mutation Feedback \\
    \hline
    \cellcolor{gray!50} OpenGBF & User Space & Multiple & Yes & Delay Injection & Branch Coverage \\ 
    \hline
    \cellcolor{gray!50} MUZZ \cite{87} & User Space & Multiple & No & Thread Priority & Thread-Aware \\ 
    \hline
    \cellcolor{gray!50} ConAFL \cite{92} & User Space & Invalid Mem. Acc. & Partial & Thread Priority & Branch Coverage \\ 
    \hline
    \cellcolor{gray!50} AutoInter-fuzzing \cite{ko2022fuzzing} & User Space & Multiple & No & Barrier/Lock & Thread-Aware \\ 
    \hline
    \cellcolor{gray!50} ConFuzz \cite{ConFuzz} & User Space & Multiple & No & None & Thread-Aware \\ 
    \hline
    \cellcolor{gray!50} Krace \cite{xu2020krace} & Kernel Space & Data Races & Yes & Delay Injection & Thread-Aware \\ 
    \hline
    \cellcolor{gray!50} Conzzer \cite{conzzer-2022} & Kernel Space & Data Races & No & Barrier/Lock & Thread-Aware \\ 
    \hline
    \end{tabular}
\caption{Taxonomy of existing state-of-the-art concurrency-aware gray-box fuzzers.}
\label{tbl:sota_fuzzers}
\end{table*}

Multiple attempts have been made to detect security vulnerabilities in concurrent programs with fuzzing ~\cite{87,92,ko2022fuzzing,ConFuzz,xu2020krace}. Here, we organize these past efforts according to five categories in the taxonomy of Table \ref{tbl:sota_fuzzers}. The first three categories concern the usability of each fuzzer: whether they apply to user programs or operating system code (\textit{Scope}), which type of bugs they are able to detect (\textit{Vulnerabilities}), and whether their code is easily accessible (\textit{Open Source}). In this regard, none of the existing state-of-the-art fuzzers satisfy our research requirements. That is, there is no fully open-source fuzzer that can detect multiple concurrency vulnerabilities in user programs. We address this gap by introducing our own concurrency-aware fuzzer in Section \ref{sec:comcurrentFuzzer}.

The last two categories concern the fuzzing techniques themselves. Specifically, the general fuzzing strategy in Algorithm \ref{alg:AFL} requires some adaptations to produce good results on concurrent programs. First and foremost, a mechanism to force the execution of a large number of different interleaving is required (\textit{Interleaving Control}). Existing fuzzers like \textit{MUZZ} \cite{87} and \textit{ConAFL} \cite{92} manipulate the thread priorities at assembly level, others like \textit{Krace} \cite{xu2020krace} inject \texttt{sleep} instruction to force a context switch, while \textit{AutoInter-fuzzing} \cite{ko2022fuzzing} and \textit{Conzzer} \cite{conzzer-2022} instrument the code with explicit synchronization barriers or thread locks. Alternatively, the interleaving exploration can be left to the natural non-determinism of the operating system like in \textit{ConFuzz} \cite{ConFuzz}. Lastly, some authors propose to change the feedback to the input mutation engine in an attempt to guide the fuzzer towards more interesting interleavings (\textit{Mutation Feedback}). We mark such attempts as \textit{Thread-Aware} as opposed to the default \textit{Branch Coverage} metrics used in sequential fuzzing. We provide more information on these state-of-the-art fuzzers in Section \ref{sec:relatedWork_fuzzing}.

\section{PROBLEM STATEMENT}
\label{sec:probleStatement}

In general, BMC and GBF tackle the problem of finding vulnerabilities in fundamentally different ways. Consequently, it is natural to ask whether combining the two techniques can lead to better coverage of the search space. More precisely, in this study, we ask the following research question:

\textbf{Research Question.} \textit{Does an ensemble of bounded model checkers and gray-box fuzzers discover more concurrency vulnerabilities and do it faster than either approach on their own?}

In addressing this question, we are confronted with many practical design challenges, the solution of which is central in the remainder of our paper:
\begin{itemize}
    
    \item \textbf{Concurrency-aware gray-box fuzzer.} As detailed in Section~\ref{sec:prelim_fuzzing}, there are some recent existing efforts to fuzz concurrent programs, but no mature open-source tool exists. Consequently, designing such a tool is an important step towards answering our research question. In doing so, we aim to draw from the lessons learned in the literature and implement \textit{OpenGBF}, a tool that is representative of state-of-the-art concurrency-aware GBF. We do so in Section~\ref{sec:comcurrentFuzzer}.
    
    \item \textbf{Aggregating BMC and GBF results.} By creating an ensemble of different tools, we run into the risk of them returning conflicting results. The main reason is that BMC relies on abstractions of the program execution states and symbolic execution (see Section \ref{sec:prelim_bmc}), whereas GBF tests concrete inputs and execution schedules. When the two approaches disagree, we have an opportunity to make an informed choice about the verification outcome. We propose to do so via a decision matrix, as detailed in Section~\ref{sec:compining}.
    
    \item \textbf{Resource allocation trade-off.} The main drawback of using an ensemble of different tools is that they all compete for the same computational resources. We must choose how many resources to allocate to each tool for applications with limited time, memory, or computational power. In general, our decisions depend not only on the problem at hand but also on the partial results we obtain from the tools in the ensemble. We discuss strategies to optimize our ensembles in Section \ref{sec:compining}.
\end{itemize}

Note that the design challenges listed above are not orthogonal. We clarify when our choices impact multiple of them in Sections~\ref{sec:comcurrentFuzzer} and~\ref{sec:compining}. Furthermore, we mention reasonable alternatives; these are left as future work.

\section{DESIGNING A STATE-OF-THE-ART CONCURRENCY GRAY-BOX FUZZER}
\label{sec:comcurrentFuzzer}

This section describes the main design challenges we address in implementing our concurrency-aware gray-box fuzzer \textit{OpenGBF}. Namely, we discuss how we control the thread interleavings (Section \ref{sec:fuzzConcurrency}) and how we generate witness information when a violation is found (Section \ref{sec:fuzzWitness}). Both of these goals require the instrumentation of the PUT as detailed in Sections \ref{sec:fuzzInstrument} and \ref{sec:Example}.

Note that our GBF is based on established techniques: fuzzer-controlled delay injection to force interleaving exploration and branch coverage to guide the fuzzer mutation engine (see Table \ref{tbl:sota_fuzzers}). At the same time, we believe that our design is worth reporting for two reasons. On the one hand, our GBF is the only user-space concurrency fuzzer that is currently available as fully open-source software; thus the present section is a useful reference for future users. On the other hand, our GBF is a transparent effort to reproduce the claims of the existing literature, which are currently impossible to confirm given the lack of open-source codebases.

\subsection{Custom LLVM Pass Instrumentation}
\label{sec:fuzzInstrument}

We build our concurrency-aware fuzzer on top of the widely used gray-box fuzzer \textit{AFL++}\cite{aflinstrumentation}, which is designed to find vulnerabilities in sequential programs. \textit{AFL++} minimizes the fuzzing overhead by instrumenting the PUT via an LLVM pass \cite{llvmpassAFL}. The LLVM pass is an essential framework of the LLVM compiler. It works with the program translated into the LLVM intermediate representation (IR) language and adds additional code to monitor the program behavior\cite{SoftwareInstrumentation}.

We combine the standard LLVM pass of \textit{AFL++} with our custom independent LLVM pass to make our fuzzer aware of concurrent execution (see Algorithm \ref{alg:ebf}). More specifically, we inject five different function calls: a delay function (see line \ref{alg:ebf:line4}), two thread-monitoring functions (see lines \ref{alg:ebf:line6} and \ref{alg:ebf:line8}) and two information-collecting functions (see lines \ref{alg:ebf:line10} and \ref{alg:ebf:line12}). The first function controls the interleaving schedule, and we explain its implementation details in Section \ref{sec:fuzzConcurrency}. The second and the third functions monitor the number of active threads (see Section \ref{sec:fuzzWitness}) in the PUT during run-time by tracking when the functions \textit{pthread\_create} and \textit{pthread\_join} are called. The last two functions record the information required to generate a witness file containing the execution trace. We present a full example of instrumented code in Section \ref{sec:Example}.

We bundle these five instrumentation functions in a run-time library. We compile and link both the runtime library and the instrumented PUT using the \textit{AFL++} clang wrapper. The resulting executable can be fuzzed to detect reachability and memory corruption bugs in the default setting. Optionally, the \textit{ThreadSanitizer} flag can be enabled for finding concurrency bugs.

\begin{algorithm}[tb]
\caption{LLVM Pass Instrumentation}\label{alg:ebf}
\textbf{Input:} $PUT$ -- program under test.\\
\textbf{Output:} $M$ -- instrumented program.\\
\textbf{Shorthands:} \\
$\lambda_d - delay\_function()$; \\
$\lambda_a - pthread\_add()$; \\ 
$\lambda_j - pthread\_release()$; \\ 
$\lambda_e - EBF\_add\_store\_pointer()$; \\
$\lambda_l - EBF\_alloca()$;
\begin{algorithmic}[1]
 \STATE $M \gets PUT$
\FORALL{Function $F \in PUT$}
\FOR{Instruction $I$ in $F$}
 \STATE $M \gets$ instrument ($\lambda_d, I, M$) \hfill\COMMENT{insert a call to $delay\_function()$ (Algorithm \ref{alg:delayfuncion}) after each instruction to run a delay at run-time}\label{alg:ebf:line4}
 \IF{$I == pthread\_create()$}
   \STATE $M \gets$ instrument ($\lambda_a, I, M$) \COMMENT{insert a call to $pthread\_add()$ (Algorithm \ref{alg:activethreadfuncion+1}) to increase the active threads counter at run-time}\label{alg:ebf:line6}
 \ELSIF{$I == pthread\_join()$}
  \STATE $M \gets$ instrument ($\lambda_j, I, M$) \COMMENT{insert a call to $pthread\_release()$ (Algorithm \ref{alg:activethreadfuncion-1}) to decrease the active threads counter at run-time}\label{alg:ebf:line8}
    \ELSIF{$I$ is DECLARATION}
  \STATE $M \gets$ instrument ($\lambda_l, I, M$) \COMMENT{insert a call to $EBF\_alloca()$ function (Algorithm \ref{alg:witnessgeneration_alloca}) to record a pair of the name and address of the variable declaration.}\label{alg:ebf:line10}
  \ELSIF{$I$ is STORE} 
  \STATE $M \gets$ instrument ($\lambda_e, I, M$) \COMMENT{insert a call to $EBF\_add\_store\_pointer()$ (Algorithm \ref{alg:witnessgenerationStore}) function to record the assignment information for witness generation}\label{alg:ebf:line12}
 \ENDIF
 \ENDFOR
    \ENDFOR
\RETURN $M$
\end{algorithmic}
\end{algorithm}

\subsection{Controlling the thread interleaving}
\label{sec:fuzzConcurrency}

As previously mentioned in Section~\ref{sec:introduction}, our main algorithmic idea is to introduce random delays in the PUT to force context switches between threads. However, there are several major corner cases that \textit{OpenGBF} needs to take care of. 

First, if the program features many active threads, we need to limit their number during the PUT execution. Limiting the number of threads is an unfortunate but necessary approximation of the PUT run-time behavior. Increasing the number of active threads slows down the PUT execution and consumes more compute resources during fuzzing. Furthermore, the PUT may attempt to create an ``infinite'' number of threads, which can either be an undefined behavior or just undecidable to solve.
We limit the number of threads by assuming that interleavings that create more threads than a pre-defined threshold are safe and start a new run with different interleavings. 
We discuss the effect of different threshold values on the bug-finding capabilities of our fuzzer in Section \ref{subsec:ebfSettings}.

Secondly, deadlocks in the PUT may cause the current interleaving to be stuck during execution. To avoid this problem, we force the fuzzer to terminate non-deterministically by introducing a probability $p$ of exiting at every instruction.

Thirdly, in \textit{EBF} we provide a mechanism for defining atomic blocks (via $EBF\_atomic\_begin$ and $EBF\_atomic\_end$ functions) within the PUT. They can be used to ensure that all instructions inside these blocks are executed atomically\footnote{This is useful since not all versions of C language provide the means for defining atomic instructions.}. To this end, our delay function will force all other threads to wait until the atomic block has finished. We achieve this by initializing a global mutex (i.e., $EBF\_mutex$) which the active thread can lock. If the global mutex is locked and the current interleaving does not own the global mutex, then we wait for the mutex owner to finish its execution.
Additionally, \textit{EBF} will not insert delays inside the atomic blocks (thus, improving the performance of the instrumented program) since no thread interleavings can take place within these blocks.

Finally, we force different interleavings by changing the amount of delay (in milliseconds) inserted after each instruction. The delay values are drawn uniformly at random from a pre-set range. More specifically, we let \textit{AFL++} produce a seed value for the random number generator providing the delay values. We explore the impact of different delay ranges on the bug-finding ability of our fuzzer in Section \ref{subsec:ebfSettings}.

The above design choices have been incorporated into the implementation of function  $delay\_function$, whose definition is illustrated in Algorithm \ref{alg:delayfuncion}. In lines \ref{alg:delayfuncion:line2}-\ref{alg:delayfuncion:line4} we implement our strategy for limiting the number of active threads. If the number of active threads $T_N$ is greater than the given threshold $T_T$ or we extract $1$ from a Bernoulli distribution with success probability $p$, then the fuzzer exits this analysis normally (see line \ref{alg:delayfuncion:line3}) and starts a new run with different interleavings (i.e., different delay values). 
Otherwise, we check whether the current thread owns the global mutex (line \ref{alg:delayfuncion:line5}), and if so, we let it finish its execution and release the mutex (lines \ref{alg:delayfuncion:line6} and \ref{alg:delayfuncion:line7}).
If the global mutex is not released before the timeout (line \ref{alg:delayfuncion:line9}), we also allow the fuzzer to exit this analysis normally (line \ref{alg:delayfuncion:line11}). This is done to prevent deadlocks if the global mutex is never released.
Finally, the delay is executed by running a \textit{sleep} function for the duration value produced by the fuzzing engine (line \ref{alg:delayfuncion:line13}).

\begin{algorithm}[tb]
\caption{Function $\_delay\_function()$}\label{alg:delayfuncion}
\textbf{Global:} $T_T$ -- thread threshold, $T_N$ -- number of threads running, $p$ -- probability of exiting, $T_C$ -- current thread, $EBF\_mutex$ -- global mutex.
\begin{algorithmic}[1]
\STATE{\textbf{Function} \_delay\_function()}

\IF{$T_N> T_T$ \OR $Bernoulli(p) == 1$}\label{alg:delayfuncion:line2}
\STATE{\bf exit}\hfill\COMMENT{exit this analysis normally}\label{alg:delayfuncion:line3}
\ENDIF\label{alg:delayfuncion:line4}

\IF{$T_C == EBF\_mutex$}\label{alg:delayfuncion:line5}
\STATE{$run\_instruction$} \hfill\COMMENT{run the current instruction}\label{alg:delayfuncion:line6}
\RETURN \label{alg:delayfuncion:line7}
\ENDIF
\STATE{$\phi \gets wait\_for\_timeout$} \hfill\COMMENT{wait until $EBF\_mutex$ is released}\label{alg:delayfuncion:line9}
\IF{$ \phi$ is timeout}
\STATE{\bf exit}\hfill\COMMENT{exit this analysis normally}\label{alg:delayfuncion:line11}
\ENDIF
\STATE {$sleep(*)$}\hfill\COMMENT{run a delay for * nanoseconds}\label{alg:delayfuncion:line13}
\STATE{\textbf{EndFunction}}
\end{algorithmic}
\end{algorithm}

Additionally, the number of active threads in the PUT is monitored by the functions $pthread\_add$ and $pthread\_release$, whose definitions are shown in Algorithms \ref{alg:activethreadfuncion+1} and \ref{alg:activethreadfuncion-1}, respectively. The former (the latter) increments (decrements) the active threads counter $T_N$ (see line \ref{alg:activethreadfuncion-1:line3}) atomically by locking the current thread (see line \ref{alg:activethreadfuncion-1:line2}) before changing the value of $T_N$ and unlocking it afterwards (see line \ref{alg:activethreadfuncion-1:line4}).

\begin{algorithm}[tb]
\caption{Function $pthread\_add()$}\label{alg:activethreadfuncion+1}
\textbf{Global:} $Mutex\_lock$, $T_N$ - active threads counter.
\begin{algorithmic}[1]
\STATE{\textbf{Function} $pthread\_add()$}
\STATE{lock thread $ \gets Mutex\_lock$}
\STATE{$T_N$++}
\STATE{unlock thread $ \gets Mutex\_lock$}
\STATE{\textbf{EndFunction}}
\end{algorithmic}
\end{algorithm}

\begin{algorithm}[tb]
\caption{Function $pthread\_release()$}\label{alg:activethreadfuncion-1}
\textbf{Global:} $Mutex\_lock$, $T_N$ - active thread counter.
\begin{algorithmic}[1]
\STATE{\textbf{Function} pthread\_release()}
\STATE{lock thread $ \gets Mutex\_lock$}\label{alg:activethreadfuncion-1:line2}
\STATE{$T_N$- -}\label{alg:activethreadfuncion-1:line3}
\STATE{unlock thread $ \gets Mutex\_lock$}\label{alg:activethreadfuncion-1:line4}
\STATE{\textbf{EndFunction}}
\end{algorithmic}
\end{algorithm}

\subsection{Witness Generation}
\label{sec:fuzzWitness}

If \textit{OpenGBF} finds a violation, we need to support the users and tools in reproducing the identified bug. To do so, we generate a crash report file with all the necessary information to reproduce the property violation. We use functions $EBF\_alloca$ and $EBF\_add\_store\_pointer$ to record all the information needed for automated witness generation: assumption values, thread ID, variable names, and function names as shown in Algorithms \ref{alg:witnessgeneration_alloca} and \ref{alg:witnessgenerationStore}. 

\begin{algorithm}[tb]
\caption{Function $EBF\_alloca()$}\label{alg:witnessgeneration_alloca}
\textbf{Inputs:} $a$ -- variable name, $f$ -- function name, $\&a$ -- variable address.\\
\textbf{Global:} $Mutex\_lock$, $witnessInfoAFL_{pid}$ -- witness file for the process with ID = $pid$.
\begin{algorithmic}[1]
\STATE{\textbf{Function} EBF\_alloca($a, f, \&a$) }
\STATE{lock thread $ \gets Mutex\_lock$}\label{alg:witnessgeneration_alloca:line2}
\STATE{$witnessInfoAFL_{pid} \gets write(a,f,\&a)$}\label{alg:witnessgeneration_alloca:line3}
\STATE{unlock thread $ \gets Mutex\_lock$}\label{alg:witnessgeneration_alloca:line4}
\STATE{\textbf{EndFunction}}
\end{algorithmic}
\end{algorithm}

\begin{algorithm}[tb]
\caption{Function $EBF\_add\_store\_pointer()$}\label{alg:witnessgenerationStore}
\textbf{Inputs:} 
$\&a$ - variable address, $l$ - line number in the code, $f$ - function name, $v$ - variable value.  \\
\textbf{Global:} $Mutex\_lock$, $witnessInfoAFL_{pid}$ -- witness file for the process with ID = $pid$.
\begin{algorithmic}[1]
\STATE{\textbf{Function} EBF\_add\_store\_pointer($\&a,l,f,v$)}
\STATE{lock thread $ \gets Mutex\_lock$}
\STATE{$witnessInfoAFL_{pid} \gets write(\&a,l,f,v)$}\label{alg:witnessgenerationStore:line3}
\STATE{unlock thread $ \gets Mutex\_lock$}
\STATE{\textbf{EndFunction}}
\end{algorithmic}
\end{algorithm}

As the fuzzing process begins, we run an initialization function before the \textit{main} method is called in the PUT. This function creates a witness file uniquely identified by the current process ID (i.e., $witnessInfoAFL_{pid}$) and sets the environment (i.e., initializes the global mutex $EBF\_mutex$, getting process id ($pid$)). Then, our custom LLVM pass inserts a function call to \textit{EBF\_alloca} (see line \ref{alg:ebf:line10} in Algorithm \ref{alg:ebf}) after each declaration instruction in the PUT, and a function call to \textit{EBF\_pointer\_add\_store\_pointer} (see line \ref{alg:ebf:line12} in Algorithm \ref{alg:ebf}) after each loading store instruction.

Algorithms \ref{alg:witnessgeneration_alloca} and \ref{alg:witnessgenerationStore} demonstrate the definitions of functions $EBF\_alloca$ and $EBF\_add\_store\_pointer$, respectively. The former records the declared variable's name, its address, and the name of the function where it has been declared in the PUT (see line \ref{alg:witnessgeneration_alloca:line3}). The latter records the assigned variable's address, the assigned value, the name of the function and the line of code where the assignment takes place in the PUT (see line \ref{alg:witnessgenerationStore:line3}). Both functions record information atomically -- the thread is locked (see line \ref{alg:witnessgeneration_alloca:line2}) before the writing occurs and unlocked afterwards (see line \ref{alg:witnessgeneration_alloca:line4}).
If the fuzzing run finishes normally (i.e., a timeout is reached or the process finishes with the exit code $0$), we delete the created witness files in a destructor function~\cite{destrctor}. 
If the fuzzer causes a crash in one of the PUT executions, we save the ID of the process that has crashed and generate a crash report by extracting the data from the witness file associated with this process ID. The resulting crash report contains the exact sequence of operations (i.e., memory accesses) that led to the PUT's crash (see Appendix \ref{app:validator} for more details). 

\subsection{Full illustrative Example}
\label{sec:Example}

\par To tie all these design choices together, we present an illustrative example. Assume that we have a concurrent PUT that has one reachability bug, as illustrated in Listing \ref{list:01}. The program contains two threads calling the same function \textit{foo} (see line \ref{list:01:line3}), which contains a loop of $5$ iterations. At the end of the execution, the value of $a$ should be $10$: Line \ref{list:01:line18} consists of a conditional statement that checks whether this is not the case, and reports an error (property violation). This error can only be reached when the reads and writes over $a$ are not correctly synchronized between the two threads.

\begin{figure*}[tb]
    \centering
    \subfloat[Mis-synchronized memory accesses.]{
        \includegraphics[width=\textwidth]{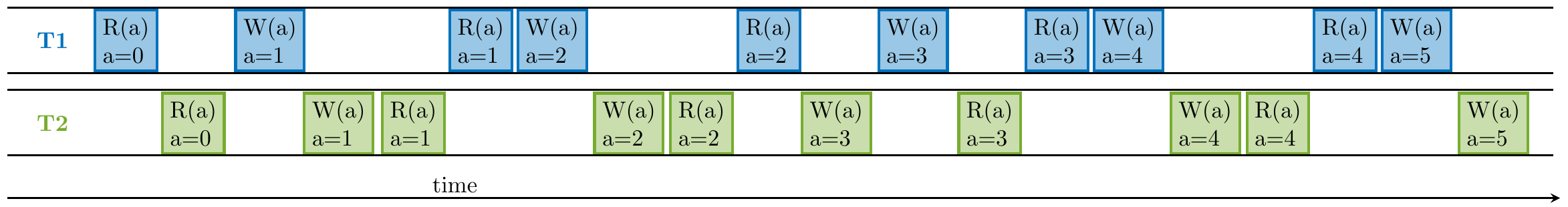}
        \label{fig:not_sync}
    } \\
    \subfloat[Synchronized memory accesses.]{
        \includegraphics[width=\textwidth]{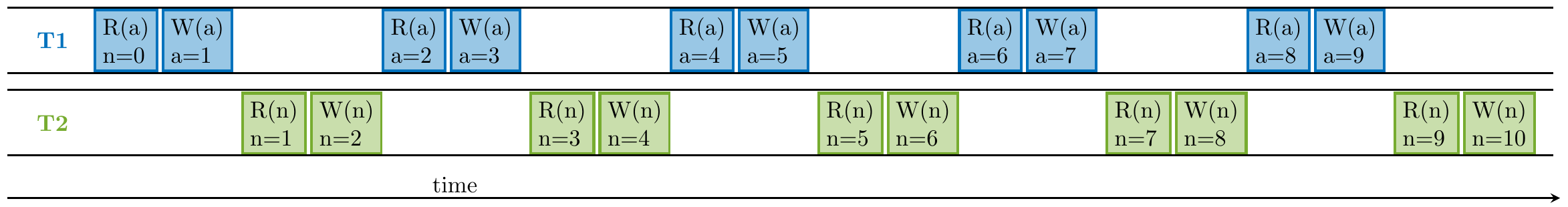}
        \label{fig:sync}
   } 
\caption{Visualization of the memory accesses on variable $a$ in Listing} \ref{list:01} for two different interleavings. In Figure \ref{fig:not_sync}, the accesses are not synchronized: both \texttt{T1} and \texttt{T2} read $a$ before simultaneously incrementing it to $a=1$. This pattern continues until the end, where the final value will be $a=5$. Conversely, Figure \ref{fig:sync} depicts synchronized accesses: \texttt{T1} reads $a$ and increments it to $a=1$, then \texttt{T2} reads $a$ and increments it to $a=2$. In the end, the final value will be $a=10$.
\end{figure*}

Figure \ref{fig:not_sync} illustrates an interleaving that causes a violation: thread 1 (T1) reads the variable $a=0$ which is initialized to $0$ in line \ref{list:01:line2}, then thread two (T2) reads $a=0$ before \texttt{T1} writes $a=1$ (see line \ref{list:01:line7}). This pattern repeats until the end, when the value of $a$ will be $5$ rather than the expected $10$.

Ideally, each thread will read the content of variable $a$ and increment it without interference from the other thread. Figure \ref{fig:sync}, illustrates an interleaving scenario where the two threads are synchronized: thread 1 (\texttt{T1}) reads $a=0$ and writes the new value $a=1$, then thread two (\texttt{T2}) reads the updated value $a=1$ and increments it to $a=2$. This pattern yields a final value of $a=10$, which makes the property hold.

When we verify the code in Listing \ref{list:01} with \textit{EBF}, we find two different bugs. Namely, \textit{OpenGBF} reports the data race, whereas all BMC tools we tested report a reachability bug at Line \ref{list:01:line18}. 

\par Let us now present an example of our instrumentation. Recall that we instrument the PUT at the LLVM-IR level. For our example of Listing \ref{list:01}, we report the LLVM-IR encoding for function \textit{foo} in Listing~\ref{list:02}. Listing~\ref{list:03} illustrates the IR after the instrumentation. In lines \ref{list:03:line7} and \ref{list:03:line20}, we call the function \textit{EBF\_add\_store\_pointer}, which is inserted after each load instruction and saves both the variable name and its value in a file to use it for generating the witness file. In lines \ref{list:03:line10},\ref{list:03:line13} and \ref{list:03:line16}, we call a function called \textit{EBF\_alloca}, which stores the metadata of any variable declared in the PUT. This information is also used to generate the witness file. In lines \ref{list:03:line11},\ref{list:03:line14},\ref{list:03:line17} and \ref{list:03:line22}, we inserting a function call to the \textit{\_delay\_function()}, as we describe in Algorithm~\ref{alg:delayfuncion}.

\begin{lstlisting}[frame=single,basicstyle=\footnotesize, language=c, style=nasm, caption=Original multi-threaded C code,label={list:01},escapechar=\%]
void reach_error() { assert(0);}
int a=0; //shared variable%\label{list:01:line2}%
void* foo(void* arg) {%\label{list:01:line3}%
 int tmp, i=1;
 while (i<=5) {
 tmp = a;%\label{list:01:line6}%
 a = tmp + 1;%\label{list:01:line7}%
 i++;%\label{list:01:line8}%
 }
 return 0;
}
int main () {
 pthread_t t1, t2;
 pthread_create(&t1, 0, foo, 0);
 pthread_create(&t2, 0, foo, 0);
 pthread_join(t1, 0);
 pthread_join(t2, 0);
 if ((a) != 10) reach_error();%\label{list:01:line18}%
} 
\end{lstlisting}

\begin{lstlisting}[frame = single, basicstyle=\footnotesize, language=llvm, style=nasm, caption=Fragment of the corresponding IR \textbf{before} instrumentation,label={list:02},escapechar=\!]
define dso_local i8* @foo(i8* %0) #0 {
  %2 = alloca i8*, align 8
  %3 = alloca i32, align 4
  %4 = alloca i32, align 4
  store i8* %0, i8** %2, align 8
  store i32 1, i32* %4, align 4
  br label %5
  %6 = load i32, i32* %4, align 4
  %7 = icmp sle i32 %6, 5
  br i1 %7, label %8, label %14                                         
  %9 = load i32, i32* @a, align 4
  store i32 %9, i32* %3, align 4
  %10 = load i32, i32* %3, align 4
  %11 = add nsw i32 %10, 1
  store i32 %11, i32* @a, align 4
  %12 = load i32, i32* %4, align 4
  %13 = add nsw i32 %12, 1
  store i32 %13, i32* %4, align 4
  br label %5
  ret i8* null
}



\end{lstlisting}

\begin{lstlisting}[frame = single, basicstyle=\footnotesize, language=llvm, style=nasm, caption=Fragment of the corresponding IR \textbf{after} instrumentation,label={list:03},escapechar=\!]
define dso_local i8* @foo(i8* %0) #0 {
  %2 = alloca i8*, align 8
  %3 = alloca i32, align 4
  %4 = alloca i32, align 4
  %5 = bitcast i8* %0 to i1*
  %6 = bitcast i8** %2 to i8*
  call void @EBF_add_store_pointer(i8* %6, i64 0, i8* getelementptr inbounds ([4 x i8], [4 x i8]* @0, i32 0, i32 0), i1* %5)!\label{list:03:line7}!
  store i8* %0, i8** %2, align 8
  %7 = bitcast i8** %2 to i8*
  call void @EBF_alloca(i8* getelementptr inbounds ([4 x i8], [4 x i8]* @1, i32 0, i32 0), i8* getelementptr inbounds ([4 x i8], [4 x i8]* @2, i32 0, i32 0), i8* %7), !\label{list:03:line10}!
  call void @_delay_function()!\label{list:03:line11}!
  %8 = bitcast i32* %3 to i8*
  call void @EBF_alloca(i8* getelementptr inbounds ([4 x i8], [4 x i8]* @3, i32 0, i32 0), i8* getelementptr inbounds ([4 x i8], [4 x i8]* @4, i32 0, i32 0), i8* %8),!\label{list:03:line13}!
  call void @_delay_function()!\label{list:03:line14}!
  %9 = bitcast i32* %4 to i8*
  call void @EBF_alloca(i8* getelementptr inbounds ([2 x i8], [2 x i8]* @5, i32 0, i32 0), i8* getelementptr inbounds ([4 x i8], [4 x i8]* @6, i32 0, i32 0), i8* %9), !\label{list:03:line16}!
  call void @_delay_function()!\label{list:03:line17}!
  %10 = sext i32 1 to i64
  %11 = bitcast i32* %4 to i8*
  call void @EBF_add_store_pointer(i8* %11, i64 6, i8* getelementptr inbounds ([4 x i8], [4 x i8]* @7, i32 0, i32 0), i64 %10), !\label{list:03:line20}!
  store i32 1, i32* %4, align 4, 
  call void @_delay_function(), !\label{list:03:line22}!
\end{lstlisting}
\section{EBF: ENSEMBLES OF BMC AND FUZZING}
\label{sec:compining}

Thanks to our \textit{OpenGBF}, we now have access to both state-of-the-art BMC and GBF tools. This section explains how we combine them in \tool and maximize their effectiveness in finding vulnerabilities in concurrent software.

We remark that types of software vulnerabilities that can be detected by \textit{EBF} solely depend on the capabilities of each individual tool used in the ensemble.
For example, most BMC tools can detect most types of illegal memory accesses (e.g., buffer overflows, use-after-free, invalid pointer dereference) and memory leaks, as well as some BMC tools, can detect concurrency bugs (i.e., thread leaks, data races, and deadlocks). Regarding \textit{OpenGBF}, its main function is exploring different executions of the program by sampling different thread schedules and different program inputs. In order to evaluate whether each such execution leads to a bug, \textit{OpenGBF} relies on the bug-detecting capabilities of sanitizers. They perform some additional instrumentation to the PUT, making it crash when a vulnerability has been detected. Namely, \textit{AddressSanitizer}~\cite{asan}is capable of identifying memory-related vulnerabilities, while \textit{ThreadSanitizer}~\cite{tsan} can identify concurrency bugs.

\begin{figure}[tb]
    \centering
    \includegraphics[width=1\columnwidth]{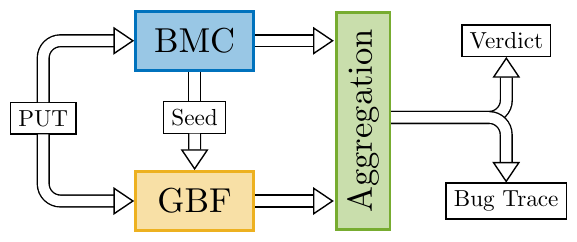}
    \vspace{-4ex}
    \caption{High-level overview of \textit{EBF}.}
    \label{fig:ensemble_pic}
\end{figure}

We present a high-level overview of the structure of our ensembles in Figure \ref{fig:ensemble_pic}. Overall, the ensemble executes both BMC and GBF tools on the PUT. The execution of these two tools is not fully independent like it would be in a portfolio \cite{XuSAT2007,BeyerCombo2022}. In fact, the result of the BMC tool run can be used to seed the GBF tool under specific conditions. We elaborate on this in Section \ref{sec:seeding}.

Furthermore, our ensemble structure in Figure \ref{fig:ensemble_pic} requires addressing the two main challenges outlined in Section \ref{sec:probleStatement}. First, the results of the BMC and GBF runs must be aggregated in a coherent assessment of the safety of the PUT. Second, the BMC and GBF tools in the ensemble compete for the same computational resources, which might reduce the ability of each tool to find violations. We present our solution to these challenges in Sections \ref{sec:ebf-aggregation} and \ref{sec:res_allocation}.

\subsection{Seeding}
\label{sec:seeding}

Before discussing our solution to the two problems of aggregation and resource sharing, let us propose a further optimization of the ensemble. Specifically, if sequential execution of the ensemble is possible, we can improve the GBF seeds by initializing them with the counterexample produced by BMC. This is only possible when the BMC tool reports a failed verification outcome. These seeds are concrete values that cause an assertion to fail. 
It is important to mention that despite the BMC tools providing thread scheduling information in their bug reports, it is not always straightforward which delay values must be applied for replicating these bug-inducing schedules.
In general, it is impossible to guarantee a particular thread order by only injecting time delays and without an explicit scheduling algorithm or another locking technique. This is because the effect of the introduced delays on the thread order depends on the implementation of multi-threading in the corresponding operating system and its current workload (e.g., sometimes, the same delay values may lead to the execution of different thread schedules). As a result, \textit{OpenGBF} uses only the bug-inducing inputs and not the thread schedule information as the seed. On the one hand, this means that \textit{OpenGBF} cannot reproduce every bug produced by the BMC tool since it might not be able to sample the sequence of delays replicating the bug-inducing thread schedule. On the other hand, this allows \textit{OpenGBF} to explore other randomly generated schedules that may lead to other bugs.


However, if the BMC engine timeouts, proves (partial) correctness or produces \textit{Unknown}, then we generate the fuzzer seed with pseudo-random integer numbers ranging between $0$ to $5000$. We determined experimentally that this range provides a good trade-off between functionality and efficiency since larger numbers (e.g., more than $5000$) 
lead the fuzzer to generate a lot of inputs that do not result in triggering a bug.
Note that these values are not directly used inside the delay function. The delays are produced by a different random number generator that is seeded from one of the inputs given by the fuzzer.

\subsection{Aggregation}
\label{sec:ebf-aggregation}

\par After running all ensemble members, we need to aggregate their outcomes. This is especially challenging since BMC and GBF may disagree on the safety of the PUT (cf.~Section~\ref{sec:probleStatement}). We summarize our aggregation rule in the decision matrix in Table~\ref{tab:aggregation}. Some decisions are straightforward: we must trust the other when either method cannot conclude. Accordingly, when GBF reports \textit{Unknown}, our decision matrix aligns with the outcome of BMC. Vice versa, when BMC cannot prove or disprove the PUT's safety, we trust the bugs found by GBF. A more interesting scenario happens when there is a conflict in the ensemble: BMC can declare a PUT as \textit{Safe}, but GBF may still be able to find a \textit{Bug}. In general, we report such instances as \textit{Conflict}. This can be caused by the over-approximations in the computational models used by the BMC tool or by the bugs from the code instrumentations introduced by the GBF tool. Each such \textit{Conflict} may be resolved by analyzing the witness file produced by the GBF.

We remark that the decision matrix proposed in Table \ref{tab:aggregation} was motivated by the SV-COMP~\cite{sv-compRules} competition rules where interactive verification is not available, 
while incorrect answers are punished by deducting competition points (see Section \ref{subsec:result1} for more details). However, verification of more complex software systems can benefit from a more descriptive decision matrix. For example, it may be useful to distinguish between different \textit{Unsafe} outcomes in Table \ref{tab:aggregation}.

\begin{table}[t]
    \centering
        \begin{tabular}{|c|c|c|c|}
        \multicolumn{4}{c}{}\\
    	\hline
    	\multicolumn{2}{|c|}{\multirow{2}{*}{EBF}} & \multicolumn{2}{c|}{\cellcolor{gray!50}GBF}\\
    	\cline{3-4}
    	\multicolumn{2}{|c|}{} & \cellcolor{gray!25}Bug & \cellcolor{gray!25}Unknown\\
    	\hline
    	\cellcolor{gray!50} & \cellcolor{gray!25}Safe & Conflict & Safe\\
    	\cline{2-4}
    	\cellcolor{gray!50} & \cellcolor{gray!25}Bug & Unsafe & Unsafe\\
    	\cline{2-4}
    	\multirow{-3}{*}{\cellcolor{gray!50}\rotatebox{90}{BMC}} & \cellcolor{gray!25}Unknown & Unsafe & Unknown\\
    	\hline
    	\multicolumn{4}{c}{}
        \end{tabular}
    \caption{\tool declares a program Safe, Unknown, Unsafe or reports a Conflict} by aggregating the outputs of \textit{BMC} and \textit{GBF}.
    \label{tab:aggregation}
\end{table}

\subsection{CPU time allocation}
\label{sec:res_allocation}

CPU time allocation is another important design choice in optimizing the performance of \tool. More specifically, we need to split the available CPU time between the two components of the ensemble in order to increase the search space coverage as much as possible for each of them. In Section~\ref{subsec:ebfSettings}, we discuss how different CPU time distribution strategies affect the overall \tool performance.

\section{EXPERIMENTAL EVALUATION}
\label{sec:results}

In this section, we demonstrate the effectiveness of BMC and GBF ensembles in a diverse set of scenarios. We will reiterate our experimental objectives before detailing the deployed benchmarks and our results.

\subsection{Objectives}

The present experimental evaluation has the following goals:

\begin{description}
   \item[EG1]\textbf{- Detection of violations in concurrent programs} \\
   Demonstrate that \tool can detect more violations in concurrent programs than state-of-the-art BMC tools on their own. 
   \item[EG2] \textbf{- Real-world performance of \textit{OpenGBF}} \\
   Demonstrate that the concurrency-aware GBF we implement in \tool can find violations in real-world programs.
   \item[EG3] \textbf{- Parameter trade-offs in our concurrency-aware fuzzer} \\
   Demonstrate that \tool produces consistent results across a wide range of parameter settings.
\end{description}

Note that the latter two objectives \textbf{EG2} and \textbf{EG3} are oriented towards demonstrating that \textit{OpenGBF} (see Section~\ref{sec:comcurrentFuzzer}) is representative of state-of-the-art gray-box fuzzing techniques.

\subsection{Results}
\label{description}

We gathered our experimental results over a substantial period, beginning in February $2021$. During this period, the design of \tool has evolved and improved. To avoid confusion, we report our results separately for each version of \tool. Namely, we start with the participation of \tool $2.3$ in the \textit{Concurrency Safety} category of SV-COMP $2022$ (see Section~\ref{subsec:result2}). This \tool version was based on \textit{CBMC} v$5.43$ and a more rudimentary implementation of our concurrency-aware fuzzer. For comparison, we also report the performance of our latest version \tool $4.0$ on the same set of benchmarks (see Section \ref{subsec:result3}). \tool $4.0$ includes the full implementation of \textit{OpenGBF} described in Section~\ref{sec:comcurrentFuzzer}, and a large number of different BMC tools. Then, we demonstrate the ability of our fuzzer to find a data race in the \textit{wolffMQTT} cryptographic library (see Section~\ref{subsec:result1}). Historically, we first found this bug in February $2021$ with an earlier version of our fuzzer. Here, we repeat our previous experiment with the latest version of \textit{OpenGBF} included in \tool $4.0$. Finally, we run an extensive comparison of the performance of \tool $4.0$ across a wide range of parameter settings (see Section~\ref{subsec:ebfSettings}).

\begin{table}[t]
    \centering
    \scalebox{0.8}{%
        \begin{tabular}{|c|c|c|c|}
        \hline
        \cellcolor{gray!50}\textbf{Verification}& \multicolumn{2}{|c|}{\cellcolor{gray!50}\textbf{Tool}} & \cellcolor{gray!50}\textbf{Score per}\\
        \cline{2-3}
        \cellcolor{gray!50}\textbf{outcome}  & \cellcolor{gray!25}\tool $2.3$& \cellcolor{gray!25}\textit{CBMC} & \cellcolor{gray!50} \textbf{benchmark}\\
        \hline
        \cellcolor{gray!25}Correct True & 139  & \textbf{148} & $\times$ 2\\
        \cline{1-4}    
        \cellcolor{gray!25}Correct False & \textbf{234} & 212 & $\times$ 1\\
        \cline{1-4}
        \cellcolor{gray!25}Correct False Unconfirmed & 55 & \textbf{90} & $\times$ 0 \\
        \cline{1-4}
        \cellcolor{gray!25}Incorrect True & 0 & 0 & $\times$ -32  \\
        \cline{1-4}
        \cellcolor{gray!25}Incorrect False & \textbf{1} & 3  & $\times$ -16 \\
        \cline{1-4}
        \cellcolor{gray!25}Unknown & 334 & 310 & $\times$ 0 \\
        \hline
        \hline
        \textit{Overall SV-COMP $2022$ score} & \textit{\textbf{496}} & \textit{460} &  \\
        \hline
        \end{tabular}}
    \caption{The results demonstrated by \tool $2.3$ and \textit{CBMC} $5.43$ in the \textit{Concurrency Safety} category of SV-COMP $2022$.
    }
    \label{tbl:SV-compResultOverall}
\end{table}

\subsubsection{\tool 2.3 participation in SV-COMP 2022}
\label{subsec:result2}

\tool $2.3$ took part in SV-COMP $2022$ in the \textit{Concurrency Safety} category \cite{svComp22concurrency}. This category features a set of $763$ concurrent C programs, $398$ of which are \textit{safe}. The bugs in the remaining $365$ programs are formulated in terms of reachability conditions: the program is deemed \textit{unsafe} if a predefined error function is reachable within the given program, and \textit{safe} otherwise. These programs contain a number of intrinsic functions \cite{sv-compRules}. We explain how we model them in Appendix \ref{app:harnessing}.

In the SV-COMP $2022$ \textit{Concurrency Safety} category, each participating tool is asked to produce one of the following six verification outcomes for a given concurrent benchmark (see the first column in Table~\ref{tbl:SV-compResultOverall}):
\begin{itemize}
    \item \textbf{Correct True.} The tool correctly confirms that the program is safe.
    \item \textbf{Correct False.} The tool correctly confirms the presence of a bug.
    \item \textbf{Correct False Unconfirmed.} The tool correctly confirms the presence of a bug, but the associated counterexample cannot be reproduced by the \textit{witness validator} tool developed by the competition organizers.
    \item \textbf{Incorrect True.} The tool confirms that a program is safe when it contains a bug.
    \item \textbf{Incorrect False.} The tool confirms that the program contains a bug when it is, in fact, safe.
    \item \textbf{Unknown.} The tool cannot conclude within the given CPU time and memory limit.
\end{itemize}

Every verification outcome is assigned a score value (see the fourth column in Table \ref{tbl:SV-compResultOverall}), which strongly discourages incorrect results. The resulting score for each tool is comprised of the sum of the scores obtained for all benchmarks.

The competition took place on the SV-COMP servers featuring $8$ CPUs (Intel Xeon E3-1230 v5 @ $3.40$ GHz) and $33$ GB of RAM. Each benchmark verification task was limited to $15$ minutes of CPU time and $15$ GB of RAM.

The version of our tool that we submitted to the competition, \tool $2.3$, is based on \textit{CBMC} v$5.43$ as a BMC engine and an earlier implementation of \textit{OpenGBF}. Namely, we selected \textit{CBMC} as it is a state-of-the-art BMC tool that has consistently been achieving high rankings in the concurrency category of SV-COMP over the past decade.  Also, the implementation of \textit{OpenGBF} we used in \textit{EBF} $2.3$ was more rudimentary. Namely, it had no limit on the number of threads, no probability of terminating early, and no mechanism to avoid injecting delays inside atomic blocks.

\tool $2.3$ reached $7$th place out of $20$ participants in SV-COMP $2022$, by scoring a total of $496$ points. Crucially, \tool $2.3$ outperformed \textit{CBMC} $5.43$, which finished $10$th with $460$ points. We report the official SV-COMP $2022$ results of these two tools in Table~\ref{tbl:SV-compResultOverall}. Note that \textit{CBMC} achieved a higher score than \tool in predicting programs safety ($148$ \textit{vs} $139$, respectively). This is an expected outcome, since \tool dedicates only $6$ minutes out of $15$ minutes to BMC, and the rest are used by \textit{OpenGBF}, which cannot prove whether a program is safe. At the same time, \tool was better than \textit{CBMC} at detecting bugs that could be confirmed by the \textit{witness validator} ($234$ \textit{vs} $212$), thus scoring extra points. 


Moreover, \tool reported only one \textit{Incorrect False} outcome, while \textit{CBMC} produced $3$ incorrect verdicts resulting in $48$ penalty points. Interestingly, \textit{EBF} avoided reproducing the latter three incorrect outcomes (returning \textit{Unknown} instead) since \textit{CBMC} did not have enough time to wrongly detect these bugs running as a part of the ensemble (reporting \textit{Unknown} as the result), while \textit{OpenGBF} also could not find any bugs in these benchmarks within the remaining time (hence, another set of \textit{Unknown}'s). In contrast, the only incorrect outcome (different from the three false positives obtained by \textit{CBMC}) produced by \tool was caused by a bug inside \textit{OpenGBF}, which made it generate a spurious counterexample.
This issue has been resolved in our most recent \tool $4.0$ version.

Overall, \tool $2.3$ improved the competition result of \textit{CBMC} $5.43$ by $\sim7.8\%$. In addition, \tool could detect and confirm a property violation in one benchmark, which could not be detected by any other dynamic tool in the competition.  These results are a first positive answer to goal \textbf{EG1}; we present further experimental evidence in Section \ref{subsec:result3}.

\subsubsection{\tool 4.0 with different state-of-the-art BMC tools}
\label{subsec:result3}

After the participation of \tool in SV-COMP $2022$, we improved \textit{OpenGBF} following the algorithmic ideas we describe in Section~\ref{sec:comcurrentFuzzer}. Here, we present the results of further experiments that test whether any BMC tool can be improved by adding our latest version of \textit{OpenGBF} on top of it (goal \textbf{EG1}). To avoid confusion, we refer to the latest implementation of our ensemble technique as \tool $4.0$.

In our evaluation, we run \tool $4.0$ over the same benchmarks from the SV-COMP $2022$ \textit{Concurrency Safety} category (see Section~\ref{subsec:result2}). However, we omit the SV-COMP aggregate scoring system (see Table~\ref{tbl:SV-compResultOverall}), since its different weights can obfuscate the advantages of each verification technique. Instead, we focus on analyzing the trade-off between proving safety\footnote{The term ``prove safety'' means that the BMC procedure could verify all reachable states and could not find an execution path that violates the safety property.} (BMC only) and bug-finding abilities (both BMC and GBF) from the raw results.

\begin{table*}[t]
    \centering
    \scalebox{1.0}{
        \begin{tabular}{|c|c|c||c|c||c|c||c|c|}
        \hline
        \cellcolor{gray!50}\textbf{Verification} &\multicolumn{8}{|c|}{\cellcolor{gray!50}\textbf{Tool}} \\
        \cline{2-9}
        \cellcolor{gray!50}\textbf{outcome} & \cellcolor{gray!25}\tool & \cellcolor{gray!25}\textit{Deagle} & \cellcolor{gray!25}\tool & \cellcolor{gray!25}\textit{Cseq} & \cellcolor{gray!25}\tool& \cellcolor{gray!25}\textit{ESBMC} & \cellcolor{gray!25}\tool& \cellcolor{gray!25}\textit{CBMC} \\
        \hline
        \cellcolor{gray!25}Correct True & 240  & 240 & 172 & \textbf{177} &65 & \textbf{70} & 139 & \textbf{146} \\
        \cline{1-9}    
        \cellcolor{gray!25}Correct False & \textbf{336} & 319 & \textbf{333} &313 & \textbf{308}& 268 & \textbf{320} & 303 \\
        \cline{1-9}
        \cellcolor{gray!25}Incorrect True & 0 & 0 & 0 & 0 & 0 & 0 & 0 &0 \\
        \cline{1-9}
        \cellcolor{gray!25}Incorrect False & 0 & 0 & 0 & 0 & 0 & 1 & 0 & 3 \\
        \cline{1-9}
        \cellcolor{gray!25}Unknown & \textbf{187} & 204 & \textbf{258} & 273 & \textbf{390} & 424 & \textbf{304} & 311 \\
        \hline
        \multicolumn{9}{c}{}
        \end{tabular}

        }
    \caption{Pair-wise comparison of the verification outcomes for \tool $4.0$ with different BMC tools ``plugged in'' against their individual performance on the benchmarks from the \textit{Concurrency Safety} category of SV-COMP 2022. 
    }
    \label{tbl:Results}
\end{table*}

Furthermore, we consider three additional BMC tools in our experiments (see Table \ref{tbl:Results}), rather than just \textit{CBMC} \cite{cbmc}. Namely, \textit{ESBMC}~\cite{78} is a powerful BMC tool that has been successfully participating in SV-COMP over the past decade. Similarly, \textit{Deagle} \cite{Deagle} and \textit{Cseq} \cite{Cseq} achieved 1st and 2nd place, respectively, in the \textit{Concurrency Safety} category at SV-COMP $2022$.

We conduct all our experiments on a virtual machine running Ubuntu 20.04 LTS with $160$ GB RAM and $25$ CPU cores of Intel Core Processor (Broadwell, IBRS) @ $2.1$ GHz. Moreover, we run \tool $4.0$ with the following parameters: maximum thread threshold $5$, delay range from $0\,[\mu s]$ to $10^5\,[\mu s]$. Additionally, we distribute the available runtime in the following way: we allocate $6$ minutes to the BMC engine, $5$ minutes to \textit{OpenGBF}, and $4$ minutes for the seeding, aggregation, and witness file generation. These parameter setting is optimal for the SV-COMP $2022$ benchmark we are using, as we discuss in Section~\ref{subsec:ebfSettings}. Note that the user can specify the time distribution between the tools in the ensemble in \tool via command-line arguments.

Table \ref{tbl:Results} reports a pair-wise comparison between \tool $4.0$ and the four different BMC tools on their own. The results demonstrate that \tool finds more bugs than all four BMC engines on their own while reducing the number of \textit{Unknown} instances. More in detail, \tool achieves the best improvement concerning \textit{ESBMC}, by finding $\sim 14.9\%$ more bugs and correcting one wrong outcome while reducing the number of safety proofs by only $\sim 7.6\%$. Similarly, the ability to double-check any counterexample produced by BMC allows \tool to correct all three erroneous outcomes produced by \textit{CBMC} while showing a marginal difference between the improvement in bug-finding ($\sim5.6\%$) and the degradation in safety proof ($\sim5\%$). In contrast, when compared to \textit{Deagle}, \tool shows no decrease in the \textit{Correct True} outcomes, but can increase the number of discovered bugs by $\sim 5.3\%$. As for \textit{Cseq}, the number of safety proofs produced by \tool declines by only $\sim 2.9\%$, while the number of \textit{Correct False} results rises by $\sim 6.3\%$.

Overall, \tool provides a better trade-off between bug-finding and safety proving than each BMC engine. On average, \tool finds over $8\%$ more concurrency bugs while reducing the number of programs declared safe by only $3.8\%$. Hence, this evaluation achieves our first experimental goal (\textbf{EG1}).

\subsubsection{Detecting a data race in \textit{wolfMQTT}}
\label{subsec:result1}

We evaluate \tool $4.0$ on the \textit{wolfMQTT} library \cite{wolfmqtt}. \textit{MQTT} (Message Queuing Telemetry Transport) is a lightweight messaging protocol developed for constrained environments like the Internet of Things (IoT). It employs the \textit{publish-subscribe} messaging pattern of publishing messages and subscribing to topics. The \textit{wolfMQTT} library is a client implementation of the \textit{MQTT} protocol written in C for embedded devices. We use its API to verify the concurrent part of the protocol implementation.

\textit{OpenGBF} detects a data race\footnote{\url{https://github.com/wolfSSL/wolfMQTT/issues/198}} in \textit{wolfMQTT} after running for $15$ minutes and consuming $24$ MB of RAM. In detail, \textit{MQTT} contains $4$ packet structures (i.e., \textit{Connect}, \textit{Publish}, \textit{Subscribe} and \textit{Unsubscribe}). The \textit{Subscribe} function accepts an acknowledgment from the server (i.e., broker). This acknowledgment was received through an unprotected pointer due to the data race detected in function \textit{MqttClient\_WaitType}, which can lead to an information leak or data corruption.
This issue has been successfully replicated and consequently fixed\footnote{\url{https://github.com/wolfSSL/wolfMQTT/pull/209}} by the \textit{wolfMQTT} developers.

Our setup for the experiment is the following. We run \tool $4.0$ on an Intel Core i7 $2.7$Ghz machine with $8$ GB of RAM running Ubuntu 18.04.5 LTS. We use a Mosquito server for the communication with the \textit{wolfMQTT} client~\cite{mosquitto}. We enable \textit{ThreadSanitizer} on top of \textit{OpenGBF} for detecting the concurrency bugs that are not formulated explicitly in terms of reaching a predefined error function (i.e., like it is done in the SV-COMP $2022$ concurrency benchmarks) or violating a safety assertion. Finally, we run our fuzzer with a thread threshold of $Th=5$ and a delay range from $0\, [\mu s]$ to $10^5\, [\mu s]$.

Other tools fail to discover the same vulnerability. More specifically, both bounded model checkers \textit{CBMC} v$5.43$ and \textit{ESBMC} v$6.8$ are unable to detect the data race within the given time limit. Moreover, the BMC tool \textit{Deagle} v$1.3$ cannot parse the program correctly, since it is using an outdated version of C parser. Similarly, \textit{Cseq} v3.0 does not support programs featuring multiple source files. Finally, neither the fuzzer \textit{AFL} nor \textit{AFL++} can detect this bug in \textit{wolfMQTT}.

As a result of this experiment, we can conclude that our second evaluation goal (\textbf{EG2}) has been achieved.

\subsubsection{Detecting memory violations in real world concurrent programs}
\label{subsec:memoryviolation}

To show scalability and robustness of \tool, we evaluate it on several real-world concurrent programs using the same machine as in Section \ref{subsec:result1}. We consider three multi-threaded real-world programs: \textit{pfscan}~\cite{pfscan}, a multi-threaded file scanner; \textit{bzip2smp}~\cite{bzip2smp}, a parallel implementation of \textit{bzip2} compressor; \textit{swarm1.1}~\cite{bader2007swarm}, a library that provides a framework for parallel programming on multi-core systems. 
Table \ref{tbl:Ral-world} presents the number of lines of code (LOC) for each PUT, the number ($N_{N}$) and types of vulnerabilities detected by \textit{EBF}, and the median time \textit{EBF} takes to find these bugs. We give more detailed information on the latter in Section \ref{subsec:fuzzer_noise}.

Both tools in \textit{EBF} detect a NULL pointer deference in \textit{pfscan} that is caused by a \texttt{malloc} instruction whose result is not checked for successful memory allocation leading to a crash due to writing to a NULL pointer. As for \textit{bzip2smp}, \textit{EBF} finds two bugs. The BMC engine detects a vulnerability in the \texttt{BZ2\_bzclose()} function, which receives a pointer that can be NULL. Meanwhile, \textit{OpenGBF} finds a memory leak in the \texttt{writerThread()} function of \textit{bzip2smp}. Regarding \textit{swarm} $1.1$, \textit{EBF} (in particular, the fuzzer) finds an invalid pointer dereference caused by an incorrect thread initialization (i.e., calling the \texttt{pthread\_create} function with a NULL pointer as an argument).

 \begin{table*}[t]
    \centering
\begin{tabular}{ |c|c|c|c|c|c|c| } 
\hline
Real-world programs & \cellcolor{gray!25} LOC&\cellcolor{gray!25}$N_{N}$ &\cellcolor{gray!25} $N_{T}$ &\cellcolor{gray!25}Median Time &\cellcolor{gray!25}ESBMC&\cellcolor{gray!25}OpenGBF\\
\hline
\cellcolor{gray!50}wolfMQTT & 9.3k & 1 & Data Race& 361.7$s$& &\checkmark\\ 
\cline{1-7}
\cellcolor{gray!50}pfscan& 1.1k & 1 &Invalid pointer dereference& 3.98$s$&\checkmark &\checkmark\\ 
\cline{1-7}
 \cellcolor{gray!50}bzip2smp & 5.3k &2 & \begin{tabular}{@{}c@{}}
                   Invalid pointer dereference\\
                   Memory leak\\
                 \end{tabular}  &
                 10.6$s$&
                 \begin{tabular}{@{}l@{}} \checkmark \\ \\
                  \end{tabular}&\begin{tabular}{@{}l@{}}  \\\checkmark \\
                  \end{tabular}\\

\cline{1-7}
 \cellcolor{gray!50}swarm 1.1 & 2.8k&1 &Invalid pointer dereference & 339.6$s$ & &\checkmark  \\ 
\cline{1-7}
\hline
\end{tabular}
    \caption{Evaluation of \textit{EBF} on real-world concurrent programs. For each program we present its size in terms of the number of lines of code (LOC), the number of vulnerabilities detected by \textit{EBF} ($N_{N}$), types of corresponding vulnerabilities ($N_{T}$), the median time (in seconds) of 20 \textit{EBF} re-runs, and which \textit{EBF} engine (i.e., ESBMC or \textit{OpenGBF}) detects the corresponding vulnerability.}
    \label{tbl:Ral-world}
\end{table*}

\subsubsection{Optimizing \tool's settings}
\label{subsec:ebfSettings}

In the following experiments, we explore different settings for \tool and \textit{OpenGBF}. For the first two experiments, we run \tool with the BMC engine switched off, allowing the fuzzer to run for $11$ minutes. While for the third evaluation, we run \tool with both engines enabled but with a different amount of time allotted (out of total $11$ minutes) to each of them.

\begin{figure}[tb]
    \centering
    \includegraphics[width=1\columnwidth]{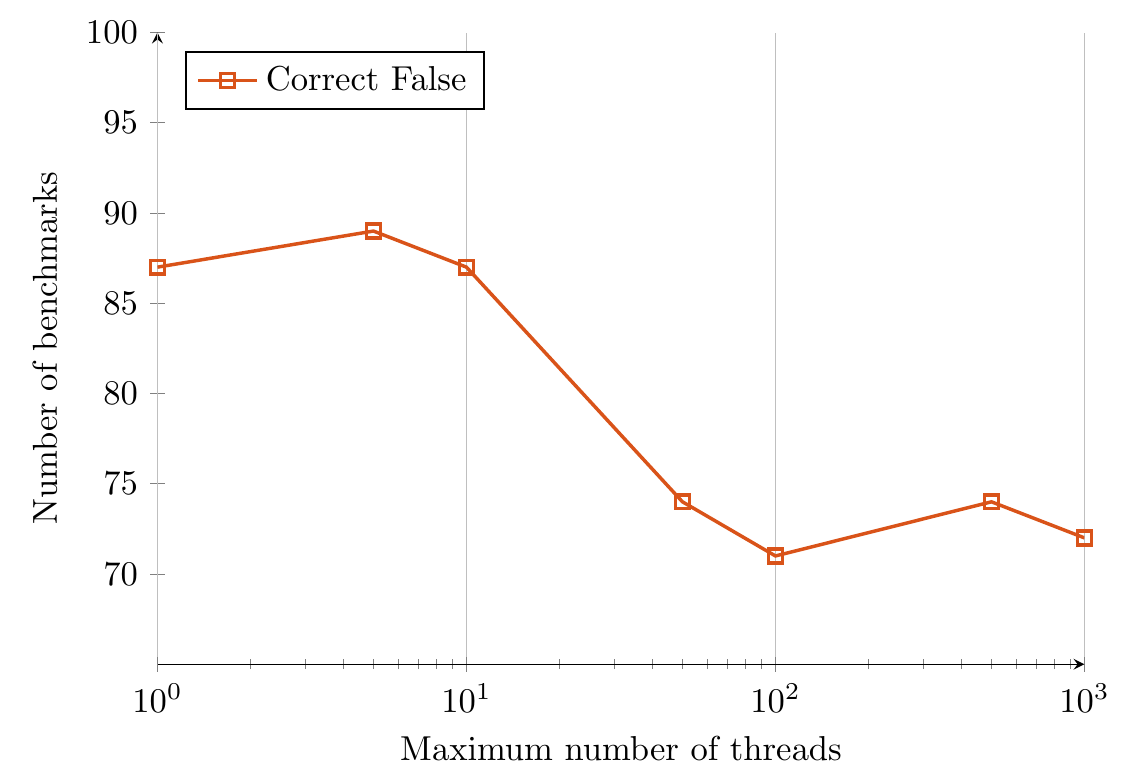}
    \vspace{-4ex}
    \caption{Number of bugs (i.e., \textit{Correct False} outcomes) discovered by \textit{OpenGBF} in \tool $4.0$ for different values of threshold on the maximum number of active threads.
    }
    \label{fig:Th1}
\end{figure}

\noindent \textbf{Maximum number of threads in \textit{\textbf{OpenGBF}}}.~
Figure~\ref{fig:Th1} shows the result of choosing different values for the thread threshold on the number of bugs (i.e., the number of \textit{Correct False} outcomes) discovered by  \textit{OpenGBF}. We ran this experiment with the delay range from $0\, [\mu s]$ to $10^5\, [\mu s]$ and probability of exiting $p=0.01\%$. It can be seen that the most optimal value lies in the region around $Th=5$, and raising the threshold value leads to fewer bugs being detected due to the increase in the number of computer resources required to maintain a more significant number of active threads. Perhaps, we can suggest that many bugs can be discovered without considering a large number of threads, which was also demonstrated by the \textit{wolfMQTT} data race that was discovered with $Th=5$. However, drawing a more robust conclusion applicable to any concurrent program requires a more extensive evaluation of our GBF on a larger set of benchmarks.

\begin{figure}[tb]
    \centering
    \includegraphics[width=1\columnwidth]{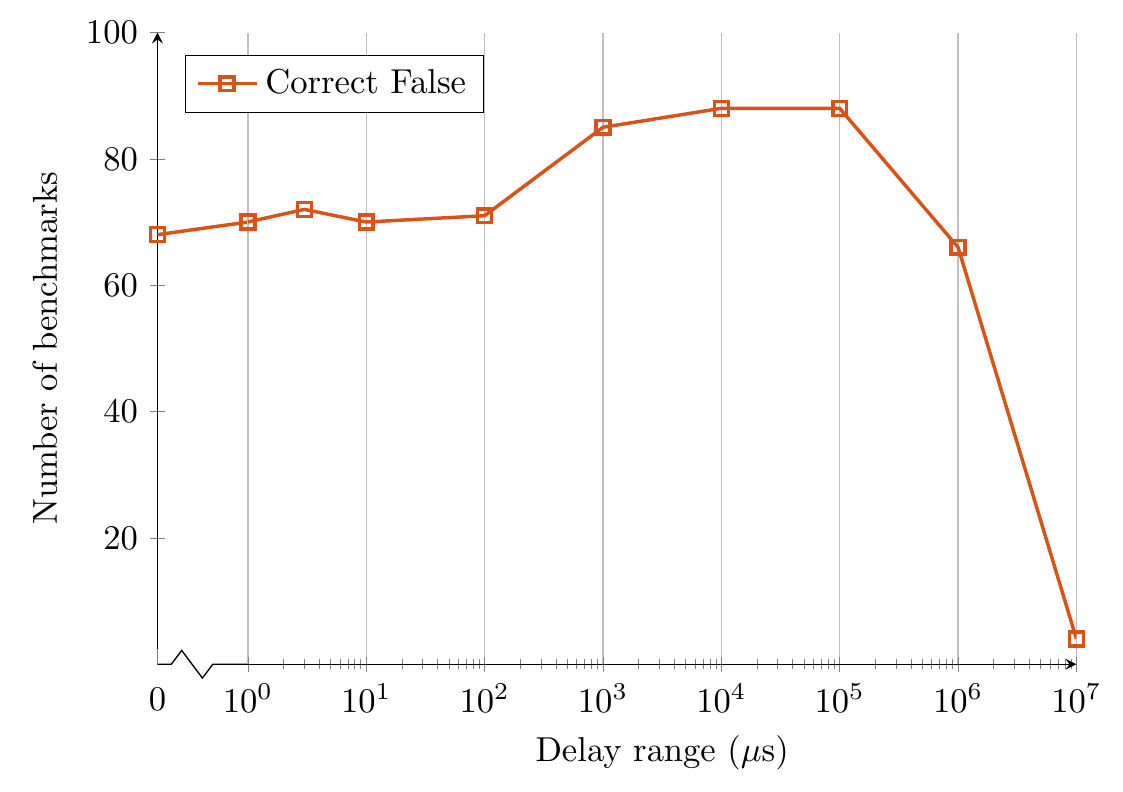}
    \vspace{-4ex}
    \caption{Number of bugs (i.e., \textit{Correct False} outcomes) discovered by \textit{OpenGBF} in \tool $4.0$ for different upper bounds of the distributions for the random delay.
    }
    \label{fig:Delays}
\end{figure}

\noindent \textbf{Maximum amount of delay in \textit{\textbf{OpenGBF}}}.~
Figure~\ref{fig:Delays} illustrates the effect of the amount of delay we insert to force scheduling in \textit{OpenGBF}. We use a logarithmic scale to compare different delay ranges in \textit{OpenGBF}. Similar to the evaluations of different thread thresholds, we use the number of \textit{Correct False} to assess the efficacy of a given delay bound. We set the thread threshold to $5$ active threads in this experiment. We change the upper bound of the delay's range from $0 \,[\mu s]$ (i.e., no delay) to $10^7\, [\mu s]$ (i.e., $10$ seconds). The results show that increasing the delay upper bound from $0$ to $10^5\, [\mu s]$ gradually improves the bug-finding capabilities of \textit{OpenGBF} from $68$ to $88$ benchmarks. When we set a large upper bound on the delay value, we increase the time range for a thread to stay inactive before it is rescheduled again, which increases the number of threads interleavings that our GBF explores. At the same time, choosing a larger upper bound (e.g., $10^6$ or $10^7$) leads to a decrease in the number of bugs found due to a higher number of timeouts. This is expected, as with larger delays the fuzzer spends the majority of the time waiting rather than executing the code. In general, we believe that finding the correct trade-off in delay range is benchmark-dependent.

\noindent
\textbf{Comparison with the ``non-instrumenting'' GBF.}
As a sanity check, we compare our GBF implementation against the ``non-instrumenting'' version of the fuzzer, which does not feature the PUT instrumentation stage described in Algorithm \ref{alg:ebf}. We run this experiment with the optimal set of parameters reported above: i.e., $Th=5$, $p=0.01\%$ and the random delay upper bound of $10^5\, [\mu s]$. The results show a nearly 50-fold increase in the number of detected bugs. More specifically, \textit{OpenGBF} detects $88$ out of $365$ vulnerabilities (i.e., $24.2 \%$ of the total), while the ``non-instrumenting'' GBF detects only $2$ out of $365$ ($0.55\%$). This highlights the necessity of using concurrency-aware fuzzers in our \tool.

\noindent \textbf{CPU time allocation inside \textit{\textbf{EBF}}.}
In this experiment we compare different ways of distributing the total verification time ($11$ minutes overall) between the fuzzer and the BMC engines in \tool.
The results demonstrate a relatively wide range of values (between $3$ and $8$ minutes per engine) within which \tool $4.0$ produces identical results detecting $320$ bugs out of $365$. At the same time, when the entire $11$ minutes are allocated to the BMC engine, the number of detected bugs drops by $\sim5\%$ to $303$ out of $365$. Conversely, the overall bug-finding performance of \tool $4.0$ decreases drastically by over $72.5 \%$ when all $11$ minutes are devoted to \textit{OpenGBF}. On the whole, this result confirms that BMC tools are better than our GBF tool when used in isolation. However, combining them both in an ensemble is going to yield better results across a very different time allocation choices.

\begin{figure}[tb]
    \centering
    \includegraphics[width=1\columnwidth]{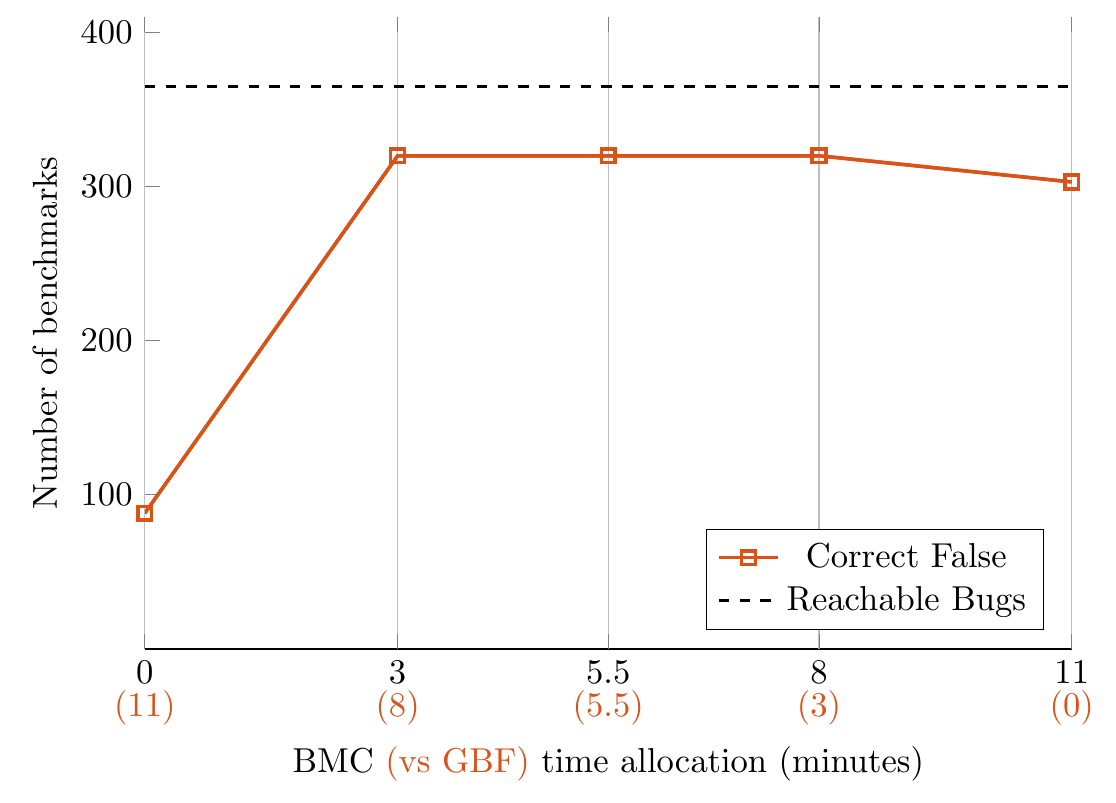}
    \vspace{-4ex}
    \caption{Number of bugs (i.e., \textit{Correct False} outcomes) discovered by \tool $4.0$ for different time allocations between the fuzzer and the BMC.}
    \label{fig:allocation}
\end{figure}

\begin{figure}[tb]
    \centering
    \includegraphics[width=1\columnwidth]{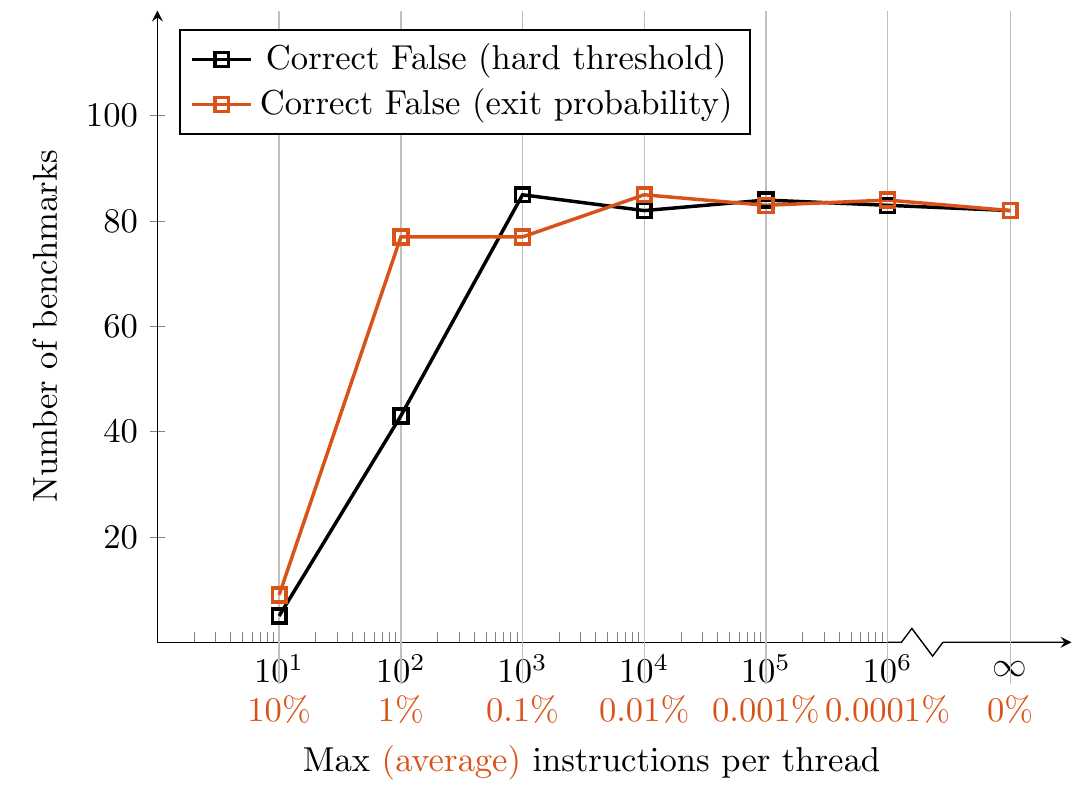}
    \vspace{-4ex}
    \caption{Number of bugs (i.e., \textit{Correct False} outcomes) discovered by \textit{OpenGBF} in \tool $4.0$ for different early thread termination strategies.}
    \label{fig:exiting}
\end{figure}

\noindent \textbf{Early thread termination in \textit{\textbf{OpenGBF}}.} Recall that our GBF fuzzer terminates the execution of each thread with probability $p$ (see Section \ref{sec:fuzzConcurrency}). This implementation detail is crucial for avoiding potential deadlocks in the PUT. Figure \ref{fig:exiting} shows the impact of different values of $p$ on the bug-finding performance of our GBF on the SV-COMP 2022 concurrency benchmark suite. For comparison, we implemented an alternative mechanism, which deterministically terminates the execution of each thread after $n$ instructions. Note that both termination mechanisms are local to each thread; thus, they do not introduce any synchronization overhead. Furthermore, we align the plots according to each thread's average number of instructions, which is the mean $1/p$ of an exponential distribution.

The results in Figure \ref{fig:exiting} show that the performance of our fuzzer is stable across a wide range of values of $p$. Interestingly, removing the termination mechanism altogether causes only minimal degradation in the fuzzer performance. Moreover, there is no significant difference between the probabilistic and deterministic termination mechanisms as the average number of instructions per thread increases. However, the performance of the probabilistic mechanism degrades more slowly as the average number of instruction decrease. We speculate that the probabilistic termination mechanism allows our GBF to explore a large number of shallow paths and a few deeper ones, thus slightly increasing the chance of finding a bug for a low average number of instructions. Finally, we select the best parameter setting $p=0.01\%$ for the rest of our experiments.

\subsubsection{Analyzing the non-determinism of our fuzzer}
\label{subsec:fuzzer_noise}

Fuzzers are fundamentally non-deterministic programs. As such, the performance of \textit{EBF} may vary across different runs. We show the effects of non-determinism by re-running our GBF 20 times on the benchmarks of the present experimental section.

\noindent \textbf{Non-determinism on SV-COMP $2022$ suite.} We run our GBF 20 times with the same optimal settings described in \ref{subsec:ebfSettings}. In the worst case, our fuzzer finds only 82 bugs, whereas in the best case, it finds 89. Given that there are 365 bugs in the SV-COMP $2022$ suite, we expect the distribution to be approximately Gaussian, with an empirical mean $85.2$ and standard deviation $2.0$. Given that the variance in the total number of bugs is small, we can trust the results of Figures \ref{fig:Th1}, \ref{fig:Delays}, \ref{fig:allocation} and \ref{fig:exiting} to give us robust values for the optimal \textit{EBF} setting.

\begin{figure}[tb]
\centering
    \includegraphics[width=1\columnwidth]{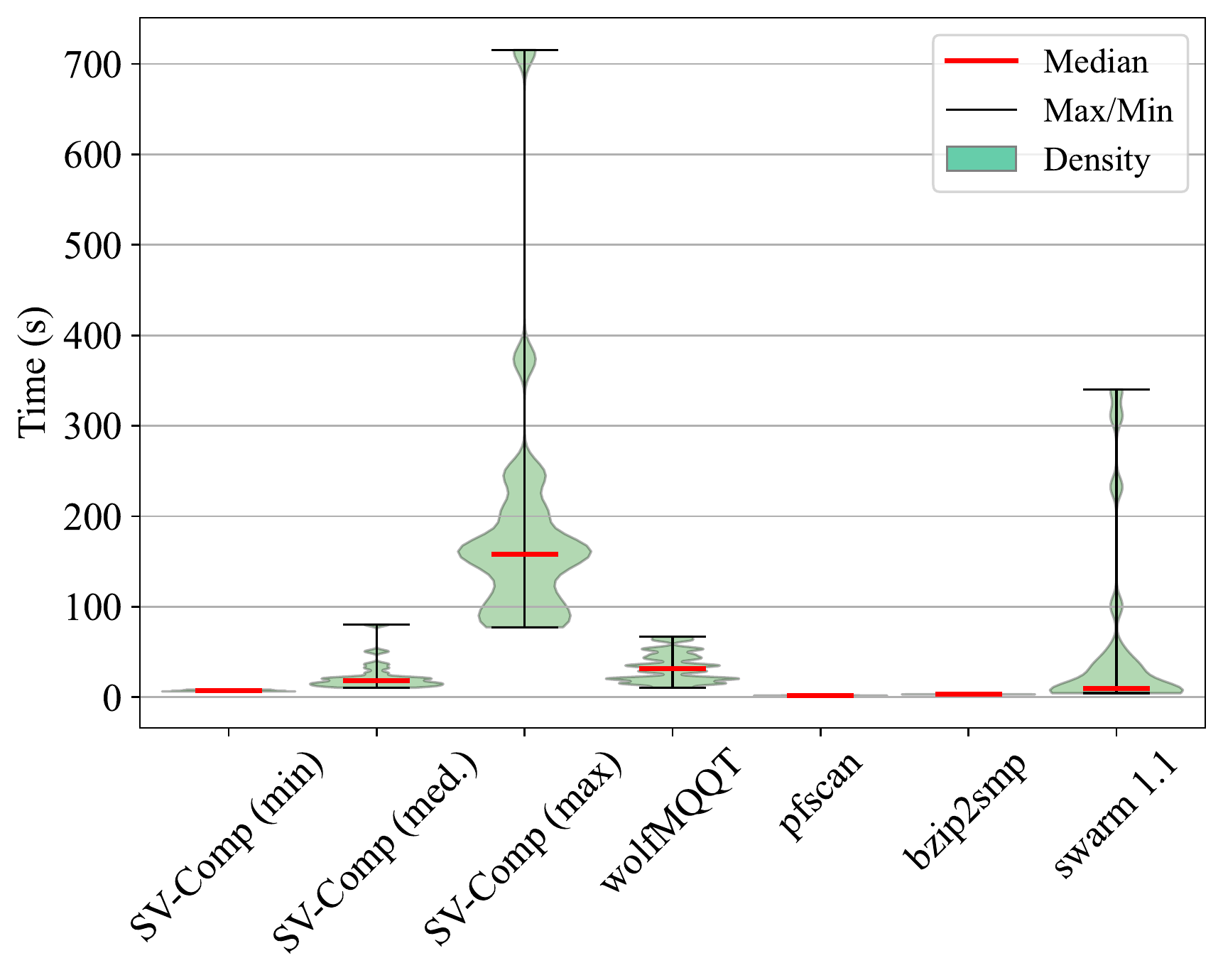}
    \vspace{-4ex}
\caption{Non-determinism of \textit{OpenGBF} across 20 re-runs of the SV-COMP'22 benchmark suite, \textit{wolfMQTT} and the real-world programs from Table \ref{tbl:Ral-world}.}
\label{fig:noise}
\end{figure}

Furthermore, we inspect the impact of fuzzer non-determinism on each individual file in the SV-COMP $2022$ benchmark suite. Specifically, there are $74$ files for which our fuzzer always finds a bug across the $20$ independent runs. Among those, we select the ones with the smallest, median, and largest variance. We plot the performance of our GBF in these three representative cases in Figure \ref{fig:noise}. Note that the violin shows the extremes of the distributions, together with their median and kernel density estimation. Since these distributions are highly non-Gaussian, we omit the mean.


\noindent \textbf{Non-determinism on \textit{\textbf{wolfMQTT}} and real-world programs.} We re-run our GBF 20 times on the \textit{wolfMQTT} library and the three real-world programs listed in Table \ref{tbl:Ral-world}. The results are shown in the violin plot of Figure \ref{fig:noise}. In the case of \textit{pfscan} and \textit{bzip2smp}, our GBF is able to find bugs almost instantly (see also Table \ref{tbl:Ral-world}). In contrast, we can observe more variance on \textit{wolfMQTT} and \textit{swarm} $1.1$. In the former case, the distribution is fairly compact in its support $[10.6s,66.8s]$. In the latter case, the distribution has a long tail. More specifically, the median time is $9.5s$, $75\%$ of the runs find a bug in less than $40s$, but there are also occasional outliers where the first bug is reported after $300s$.

\subsection{Limitations}

We identify several possible limitations to our current work.

\noindent \textbf{Incompleteness of fuzzing for proving safety.}
Fuzzing works by executing the program along many concrete paths, hoping to find the one that leads to vulnerability detection. Consequently, it cannot formally guarantee that we can exhaustively explore the entire state-space of the program. 
As a result, by design, \tool prioritizes bug-finding over proving a program's safety.

\noindent \textbf{Sources of incorrect verdicts in \textit{\textbf{EBF}}.}
Although \tool does not produce conflicting verdicts using the aggregation matrix from Table \ref{tab:aggregation}, the correctness of \tool's verification verdicts largely depends on the implementation of the tools used in the assembly.
For example, if the BMC engine produces a wrong \textit{Safe} outcome while the GBF cannot find any violations within the given time limit (thus returning \textit{Unknown}) the final verification verdict becomes \textit{Safe}.
Similarly, our GBF may become a source of an incorrect \textit{Bug} verdict when BMC reports \textit{Unknown} and the GBF crashes because of an internal bug within the GBF's implementation rather than an actual vulnerability inside the PUT. Fortunately, this is not critical since \tool generates a witness file that can be further evaluated using witness validators (see Appendix \ref{app:validator}).

\noindent \textbf{Choice of parameter settings in \textit{\textbf{EBF}}.}~
Although we conduct our evaluations over a set of more than $700$ multi-threaded C programs (see description in Section \ref{subsec:result2}), this benchmark might not represent the real-world picture of concurrent software. Thus, the optimal parameter settings for our GBF are likely to differ on another set of multi-threaded benchmarks. Nevertheless, we expect that the parameter tuning procedure on a different set of benchmarks will follow similar patterns to the ones shown in Figures \ref{fig:Th1} and \ref{fig:Delays}.

\section{RELATED WORK}
\label{sec:relatedWork}

Throughout our paper, we describe various existing studies that cover relevant tools and techniques. In this section, we collate and expand on these references. Our goal here is to clarify the context in which our research occurs.

\subsection{Bounded Model Checking (BMC)}

Bounded model checking has been successfully applied to the verification of concurrent C programs over the past years~\cite{10.1007/11513988_9}.
There exist several state-of-the-art bounded model checkers, such as \textit{ESBMC}~\cite{gadelha2020esbmc} and \textit{CBMC}~\cite{kroening2014cbmc}
that can handle both sequential and multi-threaded C programs and detect concurrency bugs (e.g., data races, deadlocks, etc.) and other vulnerabilities (e.g., buffer overflows, dangling pointers, etc.). 
In particular, \textit{ESBMC} handles concurrency by performing a depth-first search through all possible thread interleavings, up to the given maximum number of context switches~\cite{Cordeiro2012}. In contrast, \textit{CBMC} encodes each concurrent execution unit separately and joins them with partial order formulae~\cite{Alglave2013}. 

Many other BMC tools demonstrate their efficacy in the verification of concurrent programs at the annual SV-COMP software verification competition~\cite{SVCOMP21}.
For example, \textit{Lazy-CSeq}~\cite{95, 10.1007/978-3-319-08867-9_39} - one of the leaders in the concurrency category at SV-COMP over the past decade - works by translating a multi-threaded C program into a non-deterministic sequential program that considers all round-robin schedules up to a given number of rounds. Then the obtained sequential program is verified using a bounded model checker for sequential programs (e.g., \textit{CBMC}, \textit{ESBMC}). Similarly, \textit{Deagle} - the winner in the concurrency category in SV-COMP $2022$~\cite{SVCOMP22} - introduces a novel ordering consistency theory for multi-threaded programs~\cite{he2021satisfiability}, and implements a more efficient solver for this theory on top of \textit{CBMC} (front-end) and MathSAT~\cite{mathsat2008} (back-end).

\subsection{Fuzzing}
\label{sec:relatedWork_fuzzing}

Traditional techniques for fuzzing sequential programs do not translate well for concurrent programs since they let the fuzzer control only the input of the program and not the scheduling of its threads~\cite{87}. Existing proposals towards concurrency-aware fuzzing attempt to rectify this issue (see Table \ref{tbl:sota_fuzzers} for an overview).

\textit{ConAFL}~\cite{92} is a gray-box fuzzer that specializes on user-space multi-threaded programs. \textit{ConAFL} employs static analysis to locate sensitive concurrent operations to determine the execution order, focusing on three types of invalid memory access vulnerabilities: \textit{buffer-overflow}, \textit{double-free}, or \textit{use-after-free}. The thread interleavings are controlled indirectly by changing the execution priority of each thread at assembly level. As an alternative, the authors mention the possibility of injecting \texttt{sleep} commands at the code level, but they do not test it. Finally, the authors rely on the default mutation feedback of the sequential fuzzer \textit{AFL} \cite{79}, which is based on branch coverage. Due to its heavy thread-aware static and dynamic analysis, \textit{ConAFL} cannot scale to large programs. Furthermore, the authors' static analysis tool is not publicly available \cite{conAFLPersonalCommunication}.

Similarly, the gray-box fuzzer \textit{MUZZ}~\cite{87} employs static analysis to identify blocks of code that have a higher chance of triggering a concurrency vulnerability. When the code is instrumented, such blocks receive heavier instrumentation which helps the fuzzer dynamically track the execution of different schedules. To encourage the exploration of a large number of interleavings, \textit{MUZZ} manipulates the execution order by assigning random priorities to the threads at assembly level. Despite the promising experimental results, \textit{MUZZ} is not yet publicly available.

A simpler approach is implemented in the tool \textit{ConFuzz} \cite{ConFuzz}, which lets the natural non-determinism of the operating system guide the exploration of different interleavings by random chance. To compensate for that, \textit{ConFuzz} modifies the standard branch coverage feedback of the mutation engine by measuring how far each block of code is from a thread-related instruction. Seeds that execute blocks of code closer to such instructions have a higher chance of survival at each mutation. Unfortunately, the \textit{ConFuzz} tool \cite{ConFuzz} is not publicly available.


Recently, another concurrency-aware gray-box fuzzer has been proposed in \cite{ko2022fuzzing}. This tool, called \textit{AutoInter-fuzzing}, uses static analysis to identify instruction pairs that access the same memory location but are executed by different threads. Then, the program code is instrumented with synchronization barriers that control the order of execution of the instruction in each pair. Every time one such pair is encountered during regular fuzzing, the program is re-run, forcing the opposite execution order of the pair. Unfortunately, this strategy for exploring interleavings makes \textit{AutoInter-fuzzing} suffer from low path coverage compared to other fuzzers. In line with most of the fuzzers listed in the present section, \textit{AutoInter-fuzzing} is not publicly available.

\textit{Conzzer}~\cite{conzzer-2022} improves upon the ideas of \textit{AutoInter-fuzzing}. More specifically, the instruction pairs are obtained at runtime and contain information about the execution trace. The authors argue that the fuzzer can be used to explore different interleavings for a critical region by being context-aware. They also implemented their own mutation algorithm, resulting in the fuzzer being able to explore more interleaving than \textit{AutoInter-fuzzing}.

On a different note, Krace \cite{xu2020krace} is a fuzzer for kernel file systems that specializes in finding data races. We mention it here because it also employs the interleaving control strategy of injecting delays in the program code. Furthermore, it augments the standard branch coverage metrics by explicitly tracking the order of execution of any pair of instructions that access the same memory location. This feedback induces the mutation engine to explore a larger number of thread interleavings. The source code of \cite{xu2020krace} is available but cannot be used in our research as it targets data races in the kernel space.

\textit{OpenGBF} (see Section \ref{sec:comcurrentFuzzer}) implements many of these ideas, including instrumenting the code with \texttt{sleep} instructions, forcing the exploration of random interleavings and letting the fuzzer control the randomness through its mutation engine. In the future, if the aforementioned concurrency-aware fuzzers become open source \cite{EBF}, it will be possible to test their efficacy when paired with BMC tools, as we do here with our GBF tool.

\subsection{Hybrid Techniques}
\label{subsec:hybrid_techniques}

Recently, several efforts have combined fuzzing with various forms of symbolic execution and static analysis \cite{li2018fuzzing}. The rationale behind these efforts is that fuzzing alone struggles to find ``deep'' bugs and vulnerabilities because the random mutations introduced in the input have a low probability of hitting complex paths in the program. In contrast, if the fuzzer is given a set of input seeds that are already close to the correct target, the evolutionary algorithm has a higher chance of exposing the bugs and vulnerabilities.

To this end, Ognawala et al.~\cite{28} propose to increase the coverage of fuzzing by augmenting the set of input seeds with a round of concolic execution. With it, the code coverage rises significantly. There are other examples of tools employing concolic execution, such as \textit{Driller}~\cite{stephens2016driller} and \textit{QSYM}~\cite{yun2018qsym}. Similarly, Chowdhury et al.~\cite{chowdhury2019verifuzz} are concerned with the inability of off-the-shelf fuzzers to discover inputs that pass complex blocks of program logic. Their solution is using a bounded model checker to solve the corresponding reachability problem and produce concrete input seeds that satisfy the complex conditions of the program under analysis. The fuzzer is then free to explore the search space beyond that. On a different note, Alshmrany et al.~\cite{alshmrany2020fusebmc} employs a selective fuzzer if the model checker of their \textit{FuSeBMC} tool fails to find all vulnerabilities. Such fuzzer uses the statistics collected by the model checker to create a particular set of input seeds.

\tool is similar to these hybrid tools in the sense that it exploits the combined advantages of fuzzing and model checking. However, the aforementioned hybrid tools are built around a close integration between the two techniques, often requiring specific assumptions about the verification task at hand. In contrast, our ensembles are more flexible and allow virtually any existing tool to be combined together. Finally, none of the existing hybrid approaches can verify concurrent programs.

\subsection{Other techniques}
\label{subsec:others}

Other techniques for finding vulnerabilities in concurrent programs have been proposed. Wen \textit{et al.}\cite{Periodical_Scheduling} propose a controlled concurrency testing technique called \textit{Period}, which uses a periodical execution to model the execution of concurrent programs. They feed the periodical executor with a key point slice of the target program and apply an analyzer to collect feedback on runtime information. In contrast, \textit{Peahen} \cite{cai2022peahen} is an approach to combine context-sensitive and context-insensitive static techniques, namely context reduction. This context reduction consists of filtering vulnerabilities found by a context-insensitive technique with a path feasibility check. Afterward, a context-sensitive approach is used to validate the vulnerability. Finally, \textit{QL} \cite{QL} is a tool that employs reinforcement learning to guide the exploration of interleavings. This tool uses an explicit scheduler.

On a different note, there are a few methods that improve on classic verification techniques. For example, in dynamic analysis, some works focus on improving soundness and completeness \cite{cai-sound, marthur-optimal}, while other works focus on creating a new value flow analysis for interprocedural data flow that detects concurrency issues \cite{canary-dataflow}. At the same time, there are techniques that employ a different flavor of Model Checking, specifically stateless model checking (SMC)\cite{godefroid-stateless-mc}.The method was born from the intuition that caching states in Model Checking was not as effective as a stateless approach. For example, \textit{RCMC}~\cite{kokologiannakis-rcmc} and \textit{GenMC}~\cite{Kokologiannakis-genmc} rely on having a code interpreter that is able to compute a reachability graph over the program, and use system calls during the analysis to provide more accurate results.

\section{CONCLUSIONS}
\label{sec:conclusion}

Discovering vulnerabilities in concurrent programs remains a challenging problem due to the extreme explosion of the search space in the number of possible interleavings. In this paper we focus on two existing approaches to this problem: Bounded Model Checking (BMC) and Gray-Box Fuzzing (GBF). When used on their own, each approach can only find a subset of the vulnerabilities present in state-of-the-art concurrent benchmarks. Our contribution is building ensembles comprising both BMC and GBF tools, thus exploiting the complementary advantages of these two approaches. We call such ensembles \tool.

A major hindrance to the use of \tool ensembles is the current lack of mature open-source GBF tools that support concurrent testing. For this reason, we first propose our own implementation of state-of-the-art concurrency-aware fuzzing techniques, and make \textit{OpenGBF} publicly available. Then, we combine it with a large variety of state-of-the-art BMC tools, and show that the \tool ensembles so created can find up to $14.9\%$ more concurrency vulnerabilities than the BMC tools on their own. Furthermore, thanks to \textit{OpenGBF}, we are able to discover a data race vulnerability in the open-source \textit{wolfMqtt} library.

Overall, we demonstrate that \tool is an effective technique for finding vulnerabilities in concurrent programs. Still, the capability of each ensemble is directly related to the complementary qualities of its BMC and GBF building blocks. As a consequence, we believe that improving and specializing each of the two ensemble components is the most promising direction for future works. More in detail, we need faster BMC tools that rely on rougher approximations of the program under test, in order to produce a larger number of meaningful counterexamples that the GBF tool can exploit as seeds. 

\appendices
\section{Harnessing Function}
\label{app:harnessing}
Evaluating the SV-COMP $2022$ benchmarks \cite{sv-compRules} requires specific functions that must be supported by every tool participating in the competition. As a result, we model some functions for non-determinism and synchronization. The non-determinism is used to get the value of the input from the fuzzer. The synchronization is implemented using a set of functions that guarantee atomicity (i.e., to ensure no thread interleavings during a block of instructions). 
In order to make  \textit{AFL++} understand the SV-COMP specific semantics, we implement these functions as a run-time C library and link it with the benchmark at compilation time. We make the non-deterministic input functions to read the values from \textit{stdin} (i.e., standard input) when \textit{AFL++} fuzzes the PUT. To support atomicity, we rely on functions $EBF\_atomic\_begin$ and $EBF\_atomic\_end$ described in Section \ref{sec:fuzzConcurrency}.

\section{Counter example extraction}
\label{app:validator}
\tool needs to convert the crash reports discussed in Section \ref{sec:fuzzWitness} into \textit{GraphML}-based format to allow automatic witness checkers to validate the produced witness by tracking the execution path leading to the reported bug~\cite{ViolationWitnesses}. This feature of \tool is utilized in two cases: 1) when \textit{OpenGBF} reports a bug, and/or 2) when the BMC engine produces a counterexample.

\section*{Acknowledgment}

The work in this paper is partially funded by the EPSRC grants EP/T026995/1, EP/V000497/1, EU H2020 ELEGANT 957286, and Soteria project awarded by the UK Research and Innovation for the Digital Security by Design (DSbD) Programme. M. A. Mustafa is supported by the Dame Kathleen Ollerenshaw Fellowship of The University of Manchester. F. A. acknowledges the scholarship she is receiving from King Faisal University (KFU).

\section*{Copyright}
This work has been submitted to the IEEE for possible publication. Copyright may be transferred without notice, after which this version may no longer be accessible.

\bibliographystyle{IEEEtran}
\bibliography{references}

\begin{thebibliography}{10}
\providecommand{\url}[1]{#1}
\csname url@samestyle\endcsname
\providecommand{\newblock}{\relax}
\providecommand{\bibinfo}[2]{#2}
\providecommand{\BIBentrySTDinterwordspacing}{\spaceskip=0pt\relax}
\providecommand{\BIBentryALTinterwordstretchfactor}{4}
\providecommand{\BIBentryALTinterwordspacing}{\spaceskip=\fontdimen2\font plus
\BIBentryALTinterwordstretchfactor\fontdimen3\font minus
  \fontdimen4\font\relax}
\providecommand{\BIBforeignlanguage}[2]{{%
\expandafter\ifx\csname l@#1\endcsname\relax
\typeout{** WARNING: IEEEtran.bst: No hyphenation pattern has been}%
\typeout{** loaded for the language `#1'. Using the pattern for}%
\typeout{** the default language instead.}%
\else
\language=\csname l@#1\endcsname
\fi
#2}}
\providecommand{\BIBdecl}{\relax}
\BIBdecl

\bibitem{sodan2010parallelism}
A.~C. Sodan, J.~Machina, A.~Deshmeh, K.~Macnaughton, and B.~Esbaugh,
  ``Parallelism via multithreaded and multicore cpus,'' \emph{Computer},
  vol.~43, no.~3, pp. 24--32, 2010.

\bibitem{multithreadExample}
vinod, ``Multithreading realtime examples,''
  \url{https://androidmaniacom.wordpress.com/2016/12/16/multithreading-realtime-examples/},
  2022.

\bibitem{CordeiroFB20}
L.~C. Cordeiro, E.~B. de~Lima~Filho, and I.~V. de~Bessa, ``Survey on automated
  symbolic verification and its application for synthesising cyber-physical
  systems,'' \emph{{IET} Cyper-Phys. Syst.: Theory {\&} Appl.}, vol.~5, no.~1,
  pp. 1--24, 2020.

\bibitem{lu2008learning}
S.~Lu, S.~Park, E.~Seo, and Y.~Zhou, ``Learning from mistakes: a comprehensive
  study on real world concurrency bug characteristics,'' in \emph{ASPLOS},
  2008, pp. 329--339.

\bibitem{PereiraASMMFC17}
P.~A. Pereira, H.~F. Albuquerque, I.~da~Silva, H.~Marques, F.~R. Monteiro,
  R.~Ferreira, and L.~C. Cordeiro, ``Smt-based context-bounded model checking
  for {CUDA} programs,'' \emph{Concurr. Comput. Pract. Exp.}, vol.~29, no.~22,
  2017.

\bibitem{MonteiroASICF18}
F.~R. Monteiro, E.~H. da~S.~Alves, I.~da~Silva, H.~Ismail, L.~C. Cordeiro, and
  E.~B. de~Lima~Filho, ``{ESBMC-GPU} {A} context-bounded model checking tool to
  verify {CUDA} programs,'' \emph{Sci. Comput. Program.}, vol. 152, pp. 63--69,
  2018.

\bibitem{controlEngeneering}
T.~Kelly, Y.~Wang, S.~Lafortune, and S.~Mahlke, ``Eliminating concurrency bugs
  with control engineering,'' \emph{IEEE Computer}, vol.~42, pp. 52--60, 12
  2009.

\bibitem{AbstractInterpretation}
Q.~Stievenart, J.~Nicolay, W.~De~Meuter, and C.~De~Roover, ``Detecting
  concurrency bugs in higher-order programs through abstract interpretation,''
  in \emph{PPDP}, 2015, p. 232–243.

\bibitem{dwyer1994data}
M.~B. Dwyer and L.~A. Clarke, ``Data flow analysis for verifying properties of
  concurrent programs,'' \emph{SEN}, vol.~19, no.~5, pp. 62--75, 1994.

\bibitem{16}
C.~Cadar and K.~Sen, ``Symbolic execution for software testing: three decades
  later.'' \emph{Commun. ACM}, vol.~56, no.~2, pp. 82--90, 2013.

\bibitem{50}
Y.~Li, S.~Ji, C.~Lv, Y.~Chen, J.~Chen, Q.~Gu, and C.~Wu, ``V-fuzz:
  Vulnerability-oriented evolutionary fuzzing,'' \emph{CoRR}, 2019.

\bibitem{26}
M.~Aizatulin, A.~D. Gordon, and J.~J{\"{u}}rjens, ``Extracting and verifying
  cryptographic models from {C} protocol code by symbolic execution,''
  \emph{CoRR}, vol. abs/1107.1017, 2011.

\bibitem{Biere09}
A.~Biere, ``Bounded model checking,'' in \emph{Handbook of Satisfiability},
  2009, pp. 457--481.

\bibitem{gadelha2020esbmc}
M.~R. Gadelha, R.~S. Menezes, and L.~C. Cordeiro, ``{ESBMC} 6.1: automated test
  case generation using bounded model checking,'' \emph{STTT}, pp. 1--5, 2020.

\bibitem{kroening2014cbmc}
D.~Kroening and M.~Tautschnig, ``{CBMC}--c bounded model checker,'' in
  \emph{TACAS}, 2014, pp. 389--391.

\bibitem{cpachecker2011}
D.~Beyer and M.~E. Keremoglu, ``Cpachecker: A tool for configurable software
  verification,'' in \emph{CAV}, G.~Gopalakrishnan and S.~Qadeer, Eds., 2011,
  pp. 184--190.

\bibitem{he2021satisfiability}
F.~He, Z.~Sun, and H.~Fan, ``Satisfiability modulo ordering consistency theory
  for multi-threaded program verification,'' in \emph{PLDI}, 2021, pp.
  1264--1279.

\bibitem{95}
O.~Inverso, E.~Tomasco, B.~Fischer, S.~La~Torre, and G.~Parlato, ``Lazy-cseq: a
  lazy sequentialization tool for c,'' in \emph{TACAS}, 2014, pp. 398--401.

\bibitem{25}
M.~Vanhoef and F.~Piessens, ``Symbolic execution of security protocol
  implementations: Handling cryptographic primitives,'' in \emph{WOOT}, 2018.

\bibitem{69}
D.~Trabish, A.~Mattavelli, N.~Rinetzky, and C.~Cadar, ``Chopped symbolic
  execution,'' in \emph{ICSE}, 2018, pp. 350--360.

\bibitem{27}
P.~Tsankov, M.~T. Dashti, and D.~Basin, ``{SECFUZZ}: Fuzz-testing security
  protocols,'' in \emph{AST}, 2012, pp. 1--7.

\bibitem{51}
B.~S. Pak, ``Hybrid fuzz testing: Discovering software bugs via fuzzing and
  symbolic execution,'' 2012.

\bibitem{87}
H.~Chen, S.~Guo, Y.~Xue, Y.~Sui, C.~Zhang, Y.~Li, H.~Wang, and Y.~Liu,
  ``$\{$MUZZ$\}$: Thread-aware grey-box fuzzing for effective bug hunting in
  multithreaded programs,'' in \emph{SEC}, 2020, pp. 2325--2342.

\bibitem{28}
S.~Ognawala, T.~Hutzelmann, E.~Psallida, and A.~Pretschner, ``Improving
  function coverage with munch: A hybrid fuzzing and directed symbolic
  execution approach,'' in \emph{SAC}, 2018, pp. 1475--1482.

\bibitem{alshmrany2020fusebmc}
K.~M. Alshmrany, R.~S. Menezes, M.~R. Gadelha, and L.~C. Cordeiro, ``Fusebmc: A
  white-box fuzzer for finding security vulnerabilities in c programs,''
  \emph{FASE}, 2020.

\bibitem{chowdhury2019verifuzz}
A.~B. Chowdhury, R.~K. Medicherla, and R.~Venkatesh, ``Verifuzz: Program aware
  fuzzing,'' in \emph{TACAS}.\hskip 1em plus 0.5em minus 0.4em\relax Springer,
  2019, pp. 244--249.

\bibitem{XuSAT2007}
L.~Xu, F.~Hutter, H.~H. Hoos, and K.~Leyton-Brown, ``Satzilla-07: The design
  and analysis of an algorithm portfolio for sat,'' in \emph{Principles and
  Practice of Constraint Programming -- CP 2007}, C.~Bessi{\`e}re, Ed.\hskip
  1em plus 0.5em minus 0.4em\relax Berlin, Heidelberg: Springer Berlin
  Heidelberg, 2007, pp. 712--727.

\bibitem{BeyerCombo2022}
D.~Beyer, S.~Kanav, and C.~Richter, ``Construction of verifier combinations
  based on off-the-shelf verifiers,'' in \emph{Fundamental Approaches to
  Software Engineering}, E.~B. Johnsen and M.~Wimmer, Eds.\hskip 1em plus 0.5em
  minus 0.4em\relax Cham: Springer International Publishing, 2022, pp. 49--70.

\bibitem{EBF}
\url{https://github.com/fatimahkj/EBF}, 2022.

\bibitem{pfscan}
\url{https://manpages.ubuntu.com/manpages/focal/man1/pfscan.1.html}, 2022.

\bibitem{bzip2smp}
\url{http://bzip2smp.sourceforge.net/}, 2022.

\bibitem{bader2007swarm}
D.~A. Bader, V.~Kanade, and K.~Madduri, ``Swarm: A parallel programming
  framework for multicore processors,'' in \emph{IPDPS}.\hskip 1em plus 0.5em
  minus 0.4em\relax IEEE, 2007, pp. 1--8.

\bibitem{PereiraAMSCCSF16}
P.~A. Pereira, H.~F. Albuquerque, H.~Marques, I.~da~Silva, C.~B. Carvalho,
  L.~C. Cordeiro, V.~Santos, and R.~Ferreira, ``Verifying {CUDA} programs using
  smt-based context-bounded model checking,'' in \emph{SAC}, S.~Ossowski, Ed.,
  2016, pp. 1648--1653.

\bibitem{92}
C.~Liu, D.~Zou, P.~Luo, B.~B. Zhu, and H.~Jin, ``A heuristic framework to
  detect concurrency vulnerabilities,'' in \emph{ACSAC '18}, 2018, pp.
  529--541.

\bibitem{benari2006}
M.~Ben-Ari, \emph{Principles of Concurrent and Distributed Programming}, 2006.

\bibitem{satsmt2006}
R.~Nieuwenhuis, A.~Oliveras, and C.~Tinelli, ``Solving sat and sat modulo
  theories: From an abstract davis--putnam--logemann--loveland procedure to
  dpll(t),'' \emph{JACM}, vol.~53, no.~6, p. 937–977, 2006.

\bibitem{78}
\url{https://github.com/esbmc/esbmc}, 2021.

\bibitem{cbmc}
``Cbmc,'' \url{https://github.com/diffblue/cbmc}, 2022.

\bibitem{Cseq}
``Cseq,'' \url{https://www.southampton.ac.uk/~gp1y10/cseq/cseq.html}, 2022.

\bibitem{AlshmranyABC21}
K.~M. Alshmrany, M.~Aldughaim, A.~Bhayat, and L.~C. Cordeiro, ``Fusebmc: An
  energy-efficient test generator for finding security vulnerabilities in {C}
  programs,'' in \emph{TAP}, F.~Loulergue and F.~Wotawa, Eds., vol.
  12740.\hskip 1em plus 0.5em minus 0.4em\relax Sprnger, 2021, pp. 85--105.

\bibitem{lemieux2018fairfuzz}
C.~Lemieux and K.~Sen, ``Fairfuzz: A targeted mutation strategy for increasing
  greybox fuzz testing coverage,'' in \emph{ASE '18}, 2018, pp. 475--485.

\bibitem{ko2022fuzzing}
Y.~Ko, B.~Zhu, and J.~Kim, ``Fuzzing with automatically controlled
  interleavings to detect concurrency bugs,'' \emph{JSS}, p. 111379, 2022.

\bibitem{ConFuzz}
N.~Vinesh and M.~Sethumadhavan, ``Confuzz---a concurrency fuzzer,'' in
  \emph{ICTSCI e}, A.~K. Luhach, J.~A. Kosa, R.~C. Poonia, X.-Z. Gao, and
  D.~Singh, Eds., 2020, pp. 667--691.

\bibitem{xu2020krace}
M.~Xu, S.~Kashyap, H.~Zhao, and T.~Kim, ``Krace: Data race fuzzing for kernel
  file systems,'' in \emph{IEEE SP}.\hskip 1em plus 0.5em minus 0.4em\relax
  IEEE, 2020, pp. 1643--1660.

\bibitem{conzzer-2022}
Z.-M. Jiang, J.-J. Bai, K.~Lu, and S.-M. Hu, ``Context-sensitive and
  directional concurrency fuzzing for data-race detection,'' \emph{NDSS}, 2022.

\bibitem{aflinstrumentation}
\url{https://github.com/AFLplusplus/AFLplusplus/blob/stable/instrumentation/README.llvm.md/},
  2022.

\bibitem{llvmpassAFL}
``Fast llvm-based instrumentation for afl-fuzz,''
  \url{https://github.com/AFLplusplus/AFLplusplus/blob/stable/instrumentation/README.llvm.md},
  2022.

\bibitem{SoftwareInstrumentation}
T.~Kempf, K.~Karuri, and L.~Gao, \emph{Software Instrumentation}, 09 2008.

\bibitem{destrctor}
``Declaring attributes of functions,''
  \url{https://gcc.gnu.org/onlinedocs/gcc-4.7.0/gcc/Function-Attributes.html},
  2022.

\bibitem{asan}
K.~Serebryany, D.~Bruening, A.~Potapenko, and D.~Vyukov, ``Addresssanitizer: A
  fast address sanity checker,'' in \emph{USENIX}, USA, 2012, p.~28.

\bibitem{tsan}
K.~Serebryany and T.~Iskhodzhanov, ``Threadsanitizer: Data race detection in
  practice,'' in \emph{WBIA}, 2009, p. 62–71.

\bibitem{sv-compRules}
\BIBentryALTinterwordspacing
G.~D. Maayan. (2021) Sv-comp rules. [Online]. Available:
  \url{https://sv-comp.sosy-lab.org/2022/rules.php}
\BIBentrySTDinterwordspacing

\bibitem{svComp22concurrency}
\url{https://sv-comp.sosy-lab.org/2022/benchmarks.php }, 2022.

\bibitem{Deagle}
``Deagle,'' \url{https://https://github.com/thufv/Deagle}, 2022.

\bibitem{wolfmqtt}
\url{https://github.com/wolfSSL/wolfMQTT}, 2021.

\bibitem{mosquitto}
``Mosquitto,'' \url{https://mosquitto.org/}, 2021.

\bibitem{10.1007/11513988_9}
I.~Rabinovitz and O.~Grumberg, ``Bounded model checking of concurrent
  programs,'' in \emph{CAV}, K.~Etessami and S.~K. Rajamani, Eds., 2005, pp.
  82--97.

\bibitem{Cordeiro2012}
L.~Cordeiro, J.~Morse, D.~Nicole, and B.~Fischer, ``Context-bounded model
  checking with esbmc 1.17,'' in \emph{TACAS}, 2012, pp. 534--537.

\bibitem{Alglave2013}
J.~Alglave, D.~Kroening, and M.~Tautschnig, ``Partial orders for efficient
  bounded model checking of concurrent software,'' in \emph{CAV}, 2013, pp.
  141--157.

\bibitem{SVCOMP21}
D.~Beyer, ``Software verification: 10th comparative evaluation ({SV-COMP
  2021}),'' in \emph{TACAS}, 2021, pp. 401--422.

\bibitem{10.1007/978-3-319-08867-9_39}
O.~Inverso, E.~Tomasco, B.~Fischer, S.~La~Torre, and G.~Parlato, ``Bounded
  model checking of multi-threaded c programs via lazy sequentialization,'' in
  \emph{CAV}, A.~Biere and R.~Bloem, Eds., 2014, pp. 585--602.

\bibitem{SVCOMP22}
D.~Beyer, ``Progress on software verification: {SV-COMP 2022},'' in
  \emph{TACAS}, 2022.

\bibitem{mathsat2008}
R.~Bruttomesso, A.~Cimatti, A.~Franz\'{e}n, A.~Griggio, and R.~Sebastiani,
  ``The mathsat 4 smt solver,'' in \emph{CAV}, 2008, p. 299–303.

\bibitem{79}
\url{https://github.com/google/AFL}, 2021.

\bibitem{conAFLPersonalCommunication}
C.~Lie, ``personal communications,'' 2022.

\bibitem{li2018fuzzing}
J.~Li, B.~Zhao, and C.~Zhang, ``Fuzzing: a survey,'' \emph{Cybersecurity},
  vol.~1, no.~1, pp. 1--13, 2018.

\bibitem{stephens2016driller}
N.~Stephens, J.~Grosen, C.~Salls, A.~Dutcher, R.~Wang, J.~Corbetta,
  Y.~Shoshitaishvili, C.~Kruegel, and G.~Vigna, ``Driller: Augmenting fuzzing
  through selective symbolic execution.'' in \emph{NDSS}, vol.~16, 2016, pp.
  1--16.

\bibitem{yun2018qsym}
I.~Yun, S.~Lee, M.~Xu, Y.~Jang, and T.~Kim, ``$\{$QSYM$\}$: A practical
  concolic execution engine tailored for hybrid fuzzing,'' in \emph{USENIX)},
  2018, pp. 745--761.

\bibitem{Periodical_Scheduling}
C.~Wen, M.~He, B.~Wu, Z.~Xu, and S.~Qin, ``Controlled concurrency testing via
  periodical scheduling,'' in \emph{ICSE}, 2022, p. 474–486.

\bibitem{cai2022peahen}
Y.~Cai, C.~Ye, Q.~Shi, and C.~Zhang, ``Peahen: Fast and precise static deadlock
  detection via context reduction,'' 2022.

\bibitem{QL}
\BIBentryALTinterwordspacing
S.~Mukherjee, P.~Deligiannis, A.~Biswas, and A.~Lal, ``Learning-based
  controlled concurrency testing,'' \emph{PACMPL}, vol.~4, 2020. [Online].
  Available: \url{https://doi.org/10.1145/3428298}
\BIBentrySTDinterwordspacing

\bibitem{cai-sound}
\BIBentryALTinterwordspacing
Y.~Cai, H.~Yun, J.~Wang, L.~Qiao, and J.~Palsberg, ``Sound and efficient
  concurrency bug prediction,'' in \emph{ESEC/FSE}, 2021, p. 255–267.
  [Online]. Available: \url{https://doi.org/10.1145/3468264.3468549}
\BIBentrySTDinterwordspacing

\bibitem{marthur-optimal}
\BIBentryALTinterwordspacing
U.~Mathur, A.~Pavlogiannis, and M.~Viswanathan, ``Optimal prediction of
  synchronization-preserving races,'' \emph{PACMPL}, vol.~5, jan 2021.
  [Online]. Available: \url{https://doi.org/10.1145/3434317}
\BIBentrySTDinterwordspacing

\bibitem{canary-dataflow}
\BIBentryALTinterwordspacing
Y.~Cai, P.~Yao, and C.~Zhang, ``Canary: Practical static detection of
  inter-thread value-flow bugs,'' in \emph{PLDI}, 2021, p. 1126–1140.
  [Online]. Available: \url{https://doi.org/10.1145/3453483.3454099}
\BIBentrySTDinterwordspacing

\bibitem{godefroid-stateless-mc}
\BIBentryALTinterwordspacing
P.~Godefroid, ``Model checking for programming languages using verisoft,'' in
  \emph{SIGPLAN-SIGACT}.\hskip 1em plus 0.5em minus 0.4em\relax Association for
  Computing Machinery, 1997, p. 174–186. [Online]. Available:
  \url{https://doi.org/10.1145/263699.263717}
\BIBentrySTDinterwordspacing

\bibitem{kokologiannakis-rcmc}
\BIBentryALTinterwordspacing
M.~Kokologiannakis, O.~Lahav, K.~Sagonas, and V.~Vafeiadis, ``Effective
  stateless model checking for c/c++ concurrency,'' \emph{PACMPL.}, vol.~2,
  2017. [Online]. Available: \url{https://doi.org/10.1145/3158105}
\BIBentrySTDinterwordspacing

\bibitem{Kokologiannakis-genmc}
M.~Kokologiannakis and V.~Vafeiadis, ``Genmc: A model checker for weak memory
  models,'' in \emph{CAV}, A.~Silva and K.~R.~M. Leino, Eds., 2021, pp.
  427--440.

\bibitem{ViolationWitnesses}
D.~Beyer and K.~Friedberger, ``Violation witnesses and result validation for
  multi-threaded programs,''
  \url{https://www.sosy-lab.org/research/witnesses-concurrency/}, 2022.

\end{thebibliography}

\end{document}